\documentclass[12pt]{spieman}  
\usepackage{amsmath,amsfonts,amssymb}
\usepackage{graphicx}
\usepackage{setspace}
\usepackage{tocloft}
\usepackage{xcolor}
\usepackage{rotating}
\usepackage{multirow}
\usepackage{amsmath}
\usepackage{deluxetable}

\title{Incorporating Wavefront Error, Wavefront Sensing and Control, and Sensitivities into Exposure Time Calculations for Future Space Missions with the Error Budget Software (EBS)}

\author[a,*,$\ddagger$]{Sarah Steiger}
\author[b,$\dagger$,$\ddagger$]{Pin Chen}
\author[a]{Laurent Pueyo}

\affil[a]{Space Telescope Science Institute, 3700 San Martin Drive, Baltimore, MD 21218, USA}
\affil[b]{Jet Propulsion Laboratory, California Institute of Technology, 4800 Oak Grove Drive, Pasadena CA 91109, USA}

\cftpagenumbersoff{figure}
\cftpagenumbersoff{table} 
\usepackage{caption}
\begin{document} 

\maketitle

\begin{abstract}
A primary goal of NASA's Habitable Worlds Observatory (HWO) concept is to explore with high completeness the Habitable Zones (HZ) of $\sim$ 100 stellar systems and potentially acquire spectra of $\sim$ 25 terrestrial-type planets in the HZ of Sun-like stars (implying planet/star flux ratios on the order of $10^{-10}$), which places tight constraints on the performance of observatory systems. In particular, coronagraph instrumentation needs to be matured for higher throughput, deeper contrasts, and better broadband performance, while also taking into account their sensitivity and ability to mitigate the impact of telescope instability and wavefront error (WFE), which can have a profound impact on exo-Earth imaging. Due to the specific focus on the quantity of observed/characterized exo-Earths, the success of various proposed HWO mission architectures is often represented by the estimated exo-Earth candidate yield. Computation of the minimum exposure time to achieve the required signal-to-noise on a given target, using an exposure time calculator (ETC), is a key part of yield estimation. Moreover, one relies on ETCs not only for yields, but also to identify key observatory parameters that have large effects on expected exposure, and might warrant dedicated technology development.  The impacts of coronagraph sensitivity, WFE, and wavefront sensing and control (WFS\&C) have been well studied in the context of developing error budgets for missions and instruments such as the Roman Coronagraph Instrument. Nonetheless, there is currently no easily accessible and open-source way to directly incorporate the effects of these key parameters into calculating exposure times to perform trade studies for HWO. To address this gap, we developed the Error Budget Software (EBS) \textemdash an open-source tool that synthesizes sensitivity, WFE, and WFS\&C information for a variety of temporal and spatial scales, and directly interfaces with the open-source yield code EXOSIMS to produce exposure times. We also demonstrate how EBS can be used for mission error budgeting using the example of the Ultrastable Observatory Roadmap Team (USORT) observatory design. This includes a series of both single and multi-variate parameter explorations using EBS where we identify trends between raw contrast and wavefront error, and detector noise and energy resolution. We also highlight the existence of bifurcation points \textemdash points at which mission performance dramatically changes for only a moderate variation of the observatory/instrument parameter considered.  
\end{abstract}


{\noindent \footnotesize\textbf{*}Sarah Steiger,  \linkable{ssteiger@stsci.edu} }
{\noindent \footnotesize\textbf{$\dagger$}Pin Chen,  \linkable{pin.chen@jpl.nasa.gov} }

{\noindent \footnotesize\textbf{$\ddagger$}These authors contributed equally to this work.}
\begin{spacing}{2}   

\section{Introduction}
\label{sect:intro}

The \emph{Pathways to Discovery in Astronomy and Astrophysics for the 2020s} decadal survey (``Astro2020'')\cite{2021astro2020} recommends the development of a large UVOIR space telescope, now known as the Habitable Worlds Observatory (HWO)\cite{2024HWOFeinberg}. HWO will be a transformative observatory for general astrophysics and has the particularly ambitious goal of exploring with high completeness the Habitable Zones (HZ) of $\sim$ 100 star systems and potentially taking spectra of $\sim 25$ rocky planets to search for biosignatures that could be indicative of life. This will require the telescope and coronagraph system to disentangle planetary signals that are tens of billions of times fainter than the stars they orbit (planet/star flux ratios on the order of $\sim 10^{-10}$, depending on the stellar host) placing tight constraints on the performance of all observatory systems. It is for this reason that the recommendations of the Astro2020 decadal with respect to HWO include ``design[ing] the mission with ample scientific and technical margins'' and ``reduc[ing] risk by fully maturing necessary technologies prior to the mission's development phase''. This recommendation has led to the formation of multiple groups since the publishing of the report to identify key technology gaps and formulate plans to address them. In the case of exoplanet direct imaging and the coronagraph system specifically, this began with NASA's Exoplanet Exploration Program (ExEP) Coronagraph Technology Roadmap Working Group through which the following work was completed before the formulation of the Habitable Worlds Observatory Technology Maturation Project Office (TMPO). Through the work of these groups, it has become clear that not only do coronagraph instruments need to be matured for higher throughput, deeper contrasts, and better broadband performance, but that the performance of the observatory itself, specifically observatory stability, also has a profound impact on the ability to image and characterize exo-Earths. This includes wavefront errors (WFE) caused by telescope instability, WFS\&C algorithms that mitigate the effect of WFE, and individual coronagraph designs sensitivities' to WFE on a variety of spatial and temporal scales.\cite{2019CoyleULTRA}

Due to the specific focus on the quantity of observed and characterized exo-Earths included in the Astro2020 recommendations, it is unsurprising that the scientific success of various HWO mission architectures is often represented by the expected exo-Earth candidate (EEC) yield. EEC yield is the total number of expected Earth-like planets detected over the nominal lifetime of the mission as calculated by complex mission simulators and exposure time calculators (ETCs) which are packaged as yield codes. These codes, such as the Altruistic Yield Optimizer (AYO)\cite{2024HWOYields} and EXOSIMS\cite{2016SPIEDelacroix}, rely on ETCs to not only calculate yields, but also to highlight key observatory parameters that have a particularly large effect on expected exposure times and therefore should be focused on as objects of more in-depth study. 

The impact of coronagraph sensitivity, WFE, and WFS\&C has been well studied in the context of developing error budgets for missions and instruments such as the Roman Coronagraphic Instrument (Roman CGI)\cite{2020CGI} and is analytically described in numerous works\cite{2020Nemati, 2023NematiRoman, 2024NematiHWO}. Despite this, there is currently no open-source or easily accessible way to directly incorporate these key parameters into exposure time calculations to perform trade studies for HWO. To address this gap, we have developed the Error Budget Software (EBS) which synthesizes WFE, WFS\&C, and sensitivity information for a variety of temporal and spatial scales and packages it into a format which can be directly input to the EXOSIMS ETC. This allows users to directly relate these observatory and instrument parameters to the scientific output of HWO. 

In this paper, we first state our assumptions and show the derivations of key parameters in Section \ref{sec: assump_deriv}. We then describe the approach for implementing these parameters into EBS in Section \ref{sec: ebs} including a detailed description of the software implementation in Section \ref{sec: ebs_package}. We then end with a series of trade studies performed using the open-source EBS Python package in Section \ref{sec: ebs_example}. These include analyzing the relationship between WFE and raw contrast (\S \ref{sec: wfe_contrast}), and exploring the trade-off between detector noise and spectral resolution (\S \ref{sec: det_noise_r}), as well as a 23-parameter exploration of design space using EBS's Markov Chain Monte Carlo (MCMC) capability (\S \ref{sec: mcmc_example}). We then end with our discussions and conclusions in Sections \ref{sec: Discussion} and \ref{sec: Conclusions}.

\section{Assumptions and Derivations of Key Parameters}
\label{sec: assump_deriv}

To holistically derive provisional performance requirements for the coronagraph instrument in the context of the full observatory environment (including WFE, WFS\&C and sensitivity), system-level error budgets are used to allocate tolerable errors based on estimates of acceptable required integration times. For this work, we use the same methodology as that of Roman CGI outlined in Nemati et. al. 2020, 2023 \cite{2020Nemati, 2023NematiRoman}. This starts with specifying the required signal-to-noise ratio (SNR) for planet detection/characterization, as well as the Earth-equivalent planet/star flux ratio ($EEPSR = \frac{F_p}{F_s}$). For a Sun-Earth twin at 10 pc, this corresponds to $\frac{F_p}{F_s} = 2 \times 10^{-10}$ in the V-band. The total allowable flux-ratio noise ($FRN$) can then be expressed as $FRN = \frac{F_p}{F_s} / SNR$.  For instance, to detect the previously described Earth-twin with an SNR of 5 corresponds to an allowable total $FRN$ of $4 \times 10^{-11}$. 

The error-budgeting process entails allocating this total $FRN$ into various terms that correspond to both astrophysical noise sources (e.g. shot noise of residual background starlight, local zodiacal light, and exo-zodiacal light) as well as the performance levels of various observatory system elements/components, which relate to technology requirements. Once the noise allocation is set, an ETC calculates the required exposure time for planet detection/characterization. The exposure times are then assessed for acceptability based on considerations of mission lifetime and feasible observing-time allocations. This is typically repeated in an iterative process where key error terms are varied to identify break points and estimate requirements in technological performance levels. 

In the following sections, we will analytically describe how exposure times are calculated, state our assumptions, and derive how WFE, WFS\&C, and sensitivity information can be converted into a $FRN$ to be more easily included in exposure time calculations for future trade studies.        

\subsection{Time to SNR}
A useful way to compartmentalize noise is to separate it into random and systematic components. 
Without separating out the individual noise sources, we assume here that the total noise is the sum of a random term that increases linearly with time $\sigma_n^2 = r_n t $ and a systematic one that increases quadratically with time $\sigma_{\Delta I}^2 = r_{\Delta I} t^2 $. The planet signal also grows linearly with time $s_{pl} = r_{pl} t$. As a result, SNR increases with integration time for the first, random, component, and remains constant for the latter, systematic, one. Adding the two contributions together in quadrature, we can express the required integration time to achieve a given SNR as follows:

\begin{equation}
\label{eq: ttsnr}
t_{req} = \frac{r_n}{\left(\frac{r_{pl}}{SNR}\right)^2 - {r_{\Delta I}}^2} \, ,
\end{equation}

\noindent 
In practice, $r_n$ is the random noise count rate and is the sum of the photon count rate from the planet and various backgrounds. $r_{\Delta I}$ is the photon rate associated with systemic noise and fundamentally cannot be subtracted out from our measured signal (i.e. speckle residual).
This can also be thought of as the systemic noise floor and is analogous to the $CR_{NF}$ term in the formalism of Stark et. al. 2019\cite{Stark2019}. Inspection of the denominator in Eq.~\ref{eq: ttsnr} immediately reveals a fundamental breakpoint. When $r_{\Delta I} \ge r_{pl}/SNR$, no physical solution exists; the required integration time either goes to infinity or becomes negative.  We can express this condition as follows:  

\begin{equation}
\label{eq:condition}
r_{\Delta I} < \frac{r_{pl}}{SNR} \,
\end{equation}

It should be apparent that this is a critical factor in determining planet observability and mission requirements. Indeed, a finite but longer exposure time can be traded against mission lifetime whereas an infinite exposure time simply means that the observations are not possible. As will be shown in Section \ref{sec: wfe_instab}, $r_{\Delta I}$ is dependent upon on the observatory stability as well as what type of background subtraction is used and so changes depending on the implemented observing scenarios.

\subsection{Differential Imaging Observing Scenario}

In exoplanet direct imaging, a coronagraph is often used to remove as much of the on-axis light from the host star as possible in order to measure the signals from faint surrounding companions. The level of starlight suppression achieved solely by the coronagraph (called the ``raw contrast'') is typically insufficient to detect and characterize the faintest, and often most scientifically compelling, companions due to residual leaked starlight that makes its way into the final focal plane images. \cite{2011Soummer, 2023Carter, 2025AJBalmer}. This light manifests as ``speckles''  which evolve in time, can easily masquerade as faint companions, and add a source of systematic noise into the final science images. In order to remove these speckles, some form of differential imaging is typically applied. This involves developing a speckle ``reference'' image and subtracting it from the target images to produce a background-removed science image. 

The acquiring of this reference can take many forms and leverage different types of diversity to differentiate speckles and planets. Angular Differential Imaging (ADI)\cite{2006MaroisADI} requires rolling the telescope which separates the stellar speckles, which are inherent to the optical system and so will appear stationary in the image, spatially from the planet, which will appear to rotate. For Roman CGI, Reference Differential Imaging (RDI)\cite{2009RDI} will be used. Here, the telescope points to a reference star and the coronagraph instrument performs active wavefront control to create a high-contrast dark zone around the star (a process commonly referred to as ``dark-hole digging") and then acquires a reference image. The telescope then points to the target star and acquires the target image at which point the reference image can be subtracted to reduce the systematic noise and produce a background-removed science image. Not only are these techniques quite powerful in isolation, they can also be combined to provide even better subtraction and improved detection limits.\cite{2024Juillard} 

Mathematically, all of these differential imaging techniques can be described by the subtraction of a reference image ($|E_{r}(u, v)|^2$) from a target image ($|E_{t}(u, v)|^2$). Importantly, if the telescope system is not perfectly stable between the acquisition of the reference and target images then this will manifest as a change in the wavefront $\Delta E(u, v)$. We can write this as:


\begin{equation}
\label{eq:ref_sub}
\begin{split}
|E_f(u, v)|^2 &= |E_{t}(u, v)|^2 - |E_{r}(u, v)|^2 \\
&= |E_{r}(u, v) + \Delta E(u, v)|^2 - |E_{r}(u, v)|^2 \\
&= |\Delta E(u, v)|^2 + 2\mathrm{Re}\{E_r(u, v)^* \Delta E(u, v)\}
\end{split}
\end{equation}

\noindent where $E_f(u, v)$ is the electric field constituting the background-subtracted image, and $u$, $v$ are image-plane coordinates. Note that $|E_r|^2$ is the raw contrast map and $E_r(u, v)^*$ denotes the complex conjugate of the reference electric field. As is evident in Eq.~\ref{eq:ref_sub}, wavefront changes during the time between the reference and target acquisitions produce a systematic noise that cannot be mitigated via integration. Spatial heterodyning between these wavefront changes and the static speckles (associated with raw contrast) produce a systematic noise which contributes to the total speckle residual noise ($r_{\Delta I}$) and will be parameterized in the following subsection. This effect is often referred to as the ``cross term'', ``mixing term'', or ``heterodyne term''. 

In addition to more ``traditional'' differential imaging techniques for exoplanet imaging such as ADI and RDI, this formalism (and the extension of this formalism in the following sections) is applicable to any observing scenario where (1) reference image subtraction is used and (2) the speckle field changes between the reference image and the target image. This includes newer techniques being explored for Roman CGI and HWO such as Coherent Differential Imaging (CDI)\cite{2017BottomCDI,2024RedmondCDI}. In the case of CDI however, the reference acquisitions are much more temporally proximate to the target acquisitions and so the timescales of the wavefront aberrations that it is sensitive to are much shorter.

\subsection{Modeling the Effects of WFE Instabilities}
\label{sec: wfe_instab}

Following  the formalism in Nemati et. al. 2020\cite{2020Nemati}, we write the reduction in speckle noise due to differential imaging as the ratio of the spatial standard deviation of the post-subtraction differential contrast map, $\sigma_{\Delta I}$, over the mean counts of the raw speckles, $r_{sp} t$:

\begin{equation}
    f_{\Delta I}\equiv \frac{\sigma_{\Delta I}}{r_{sp}t}
    \label{eq:f_delta_I}
\end{equation}

This ratio $f_{\Delta I}$ contains the combined contribution of two factors encoded in $\sigma_{\Delta I}$: the contrast stability  $f_{\Delta C} = \frac{\sigma_{\Delta C}}{\bar{C}}$ and the post-processing gain $f_{pp}$. Here $\bar{C}$ is the spatially averaged raw contrast. A higher $f_{\Delta C}$ indicates that the speckle field is more stable and thus can be more effectively subtracted in differential imaging and a higher $f_{pp}$ indicates an additional improvement from post-processing algorithms beyond simple subtraction. Including these terms and doing some rearranging of Eq.~\ref{eq:f_delta_I}, we can write an expression for the image residual standard deviation $\sigma_{\Delta I}$ and the rate of residual speckles as:


\begin{eqnarray}
\sigma_{\Delta I} &=& f_{\Delta C}f_{pp}r_{sp} t = r_{\Delta I} t  \\
r_{\Delta I} &=&  f_{\Delta C}f_{pp}r_{sp}
\label{eq:sigma_delta_I}
\end{eqnarray}

We remind the reader that $r_{\Delta I} $ is key driving parameter for time to SNR in Eq.~\ref{eq: ttsnr}: it depends on the raw speckle count rate, the intrinsic speckle stability, and a possible post-processing factor beyond simple subtraction. Nemati et. al. 2023\cite{2023NematiRoman} derives the relationship between the $\Delta E$ in Eq.~\ref{eq:ref_sub} and the contrast stability $\sigma_{\Delta I}$, and thus $\sigma_{\Delta C}$ up to a normalization factor. Next we use that expression to relate $r_{\Delta I} $ to the actual statistical properties of the input wavefront.







\subsection{Deriving the Speckle Stability ($f_{\Delta C}$)}

Now, we will derive how WFE, WFS\&C, and coronagraph sensitivity information can be incorporated into the speckle stability defined by Eq.~\ref{eq:sigma_delta_I} and ultimately the time to SNR given by Eq.~\ref{eq: ttsnr}. In this paper, we work under the assumption that the heterodyne term in Eq.~\ref{eq:ref_sub} dominates the contrast stability budget, which we note may not hold over large ranges of WFE instability and contrast. We first consider the case of a single spatial and temporal scale for the WFE. Then, from Nemati et. al. 2023\cite{2023NematiRoman} Eq.~70, we get:

\begin{equation}
\sigma_{\Delta C}^2  =  \bar{C} \sigma_{\Delta E}^2
\end{equation}

\noindent where $\sigma_{\Delta E}$ is the variance of the normalized (in units of contrast) focal plane electric fields associated with the spatio-temporal scales considered. We write the variance of the wavefront errors as $\sigma_{WFE}^2 = \delta^2 $: the intrinsic coronagraph sensitivity $\beta$ then relates E-field variance and  wavefront variance: $\sigma_{\Delta E} = \beta \delta$. In practice, these sensitivities are calculated by taking the partial derivatives of the normalized intensities with respect to the wavefront around a given reference contrast $C_0$:
\begin{equation}
   s =  \frac{\partial {\bar{C}}}{\partial \delta} \Big|_{C_0}.
\end{equation}
Under our assumption that the heterodyne terms dominates, Eq.~\ref{eq:ref_sub} yields $s^2 = C_0 \beta^2$. As a result, we find:
\begin{equation}
\label{eq:sigma_delta_c}
\sigma_{\Delta C}^2  =  \frac{\bar{C}}{C_0} s^2 \delta^2
\end{equation}

In EBS, the WFEs in the system at various spatial and temporal scales increase the speckle residual noise (which can also be thought of as a degradation in contrast) which is typically azimuthally averaged and parameterized at a given angular separation. WFS\&C corrects these WFEs in real time and so acts as a damping factor, diminishing the realized contrast degradation. Different coronagraph technologies then have varying sensitivities to WFEs at a given temporal and spatial frequency and so an additional amplification or mitigation factor is applied to the speckle residual noise to account for that.

Mathematically, we write the residual WFE with respect to  $i$ temporal and $j$ spatial modes after WFS\&C ($r_{ij}$) as

\begin{equation}
\label{eq: r_ij}
r_{ij} = \gamma_{ij} \delta_{ij}
\end{equation}

\noindent where $\gamma_{ij}$ are ``instability-transmission factors'' representing the performance of a WFS\&C subsystem in the coronagraph instrument. $\gamma_{ij}$ can take values between 0 and 1 where a smaller value represents better damping.  $\delta_{ij}$ is the wavefront error at the coronagraph's entrance pupil (in pm). 




We define the spatially averaged contrast at a given angular coordinate $k$ (for a given spatial mode) as $\bar{C}_k$, with an associated sensitivity $s_{kj}$. $s_{kj}$ is in units of parts per trillion per picometer (ppt/pm) and is the relationship between how much the contrast will change for a given WFE at the coronagraph's entrance pupil as a function of angular separation $k$ and spatial mode $j$.

From this we can derive the change in contrast $\sigma_{\Delta C_{k=k'}}$ (Eq.~\ref{eq:sigma_delta_c}) expected in a given angular band by taking the root mean squared standard deviation of the sensitivities multiplied by the WFS\&C instability-transmission factors. 

\begin{equation}
\label{eq: deltacon}
\begin{split}
\sigma_{\Delta C_{k=k'}} &= 10^{-12} \sqrt{\frac{C_{k'}}{C_0}} 
\sqrt{\sum_j{\left(s_{k'j}^2\sum_i{r_{ij}^2}\right)}} \\
&= 10^{-12} \sqrt{\frac{C_{k'}}{C_0}} \sqrt{\sum_i\sum_j{\left(s_{k'j}^2{r_{ij}^2}\right)}}\\
&= 10^{-12}\sqrt{\frac{C_{k'}}{C_0}} \sqrt{\sum_i\sum_j{\left(s_{k'j}{r_{ij}}\right)^2}}
\end{split}
\end{equation} 


We can now write out the contrast stability ( $f_{\Delta C} = \frac{\sigma_{\Delta C}}{\bar{C}}$) at a given angular separation ($k'$) as: 

\begin{equation}  
\label{eq: f_delta_ck_final}
\begin{split}
    f_{\Delta C_{k'}} &= \min\left\{ 1, \frac{\sigma_{\Delta C_{k'}} }{\bar{C}_{k'}}\right\} \\
\end{split}
\end{equation}  

\section{The Error Budget Software}
\label{sec: ebs}
As previously stated, deriving provisional requirements using an error budget involves an iterative process of varying WFE instability and coronagraph performance parameter values, and assessing the resulting $t_{req}$ to identify break points using the equations outlined in the previous section. Open-source yield calculation codes like EXOSIMS\cite{2016SPIEDelacroix} however, do not have the ability to incorporate this spatial and temporal WFE instability information encoded in Eqs.~\ref{eq: deltacon} and \ref{eq: f_delta_ck_final} into their calculations. This highlights the need for an open-source tool that not only performs these calculations, but also easily interfaces with existing yield codes and ETCs to both take advantage of their existing features, and to enable in-depth analyses and trade studies that include telescope instability. 

To this end, we developed the Error Budget Software (EBS)\footnote{https://github.com/chen-pin/ebs}. At the highest level, EBS is a Python wrapper around EXOSIMS which it uses to calculate exposure times and adds functionality to account for the degradation in expected contrast due to introduced wavefront instabilities. This takes into account the damping effect of WFS\&C systems, and the varying sensitivities coronagraph systems have to specific spatial and temporal modes. For this work specifically, we will be using the spatial and temporal modes defined by the Ultra-stable Large Telescope Research and Analysis (ULTRA) Program\cite{2019CoyleULTRA} summarized in Tables \ref{table:temp_mode_def} and \ref{table:spatial_mode_def}.

EBS also contains features to easily explore a range of parameter spaces in exposure time without running full yield code simulations. This includes a command line interface to sweep through values of a single observatory parameter and plot expected exposure times, as well as MCMC and Nested Sampling sweep tools to perform multivariate analyses. 

\begin{figure}
    \centering    
    \includegraphics[width=0.9\linewidth]{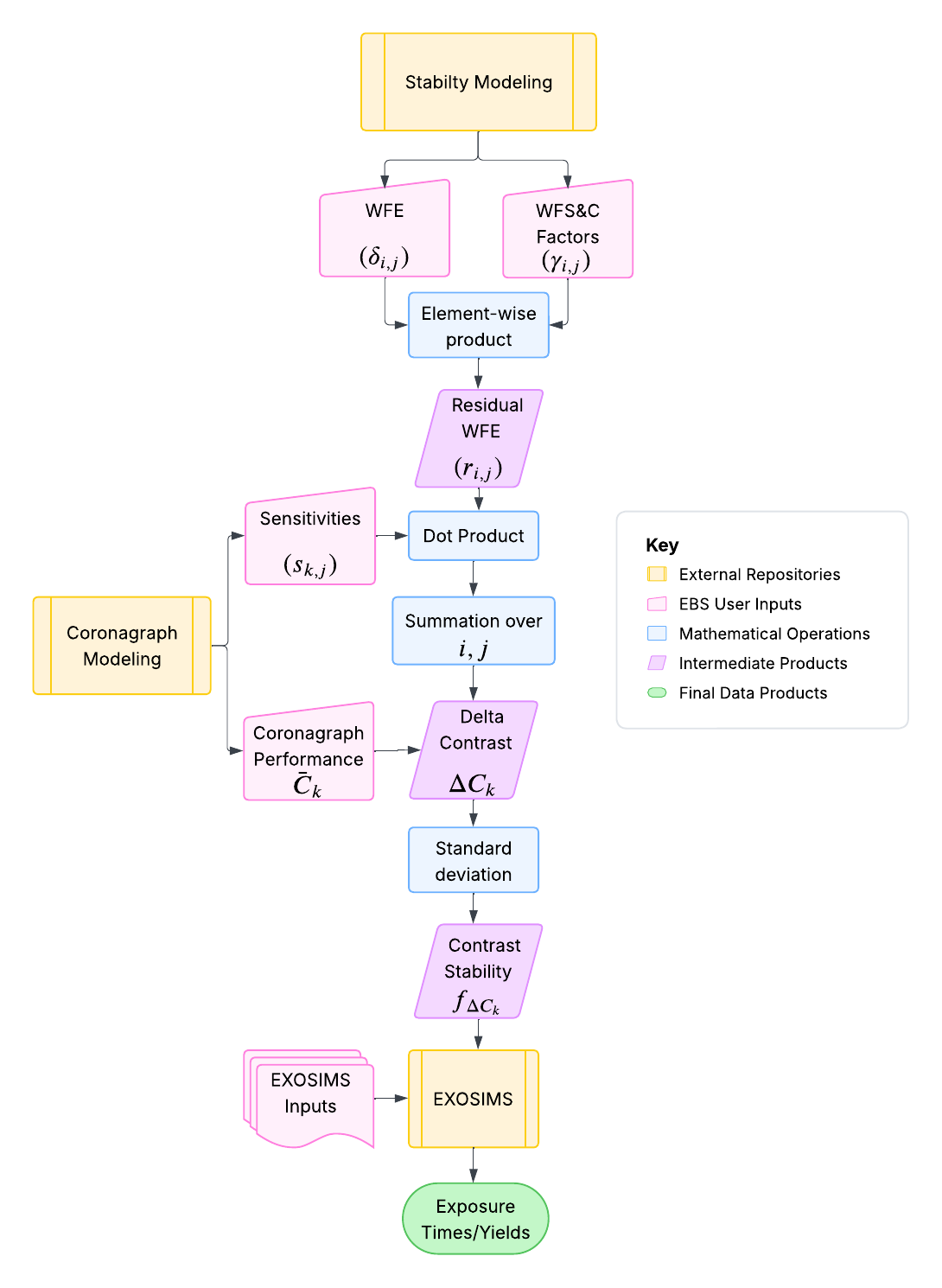}
    \caption{Flowchart depicting the basic operations of EBS as described in Section \ref{sec:core_func}.}
    \label{fig:flowchart}
\end{figure}

\subsection{Core Functionality}
\label{sec:core_func}
EBS works by calculating $f_{\Delta C}$ as defined in Eq.~\ref{eq: f_delta_ck_final} given the following 2D input arrays: 

\begin{enumerate}
    \item \textit{WFE instability ($\delta_{ij}$)}:  A parametric representation of WFE instabilities incident on the coronagraph's entrance pupil. This is a 2D array of temporal ($i$) and spatial ($j$) frequency bins and represents the changes in WFE during the time lapse between reference and target images for a differential imaging observing scenario. For each bin, the user inputs values representing RMS WFE-variance (in pm).  Mathematically, each value corresponds to the square root of a 2D spatial-temporal integration over the WFE power spectral density (PSD), bounded by the frequency limits of the frequency bins. At this stage, only notional estimates are attainable for HWO, though they will continue to be refined as the mission progresses. 
    \item \textit{WFS\&C instability-transmission factor ($\gamma_{ij}$)}: A parametric representation of the performance of active wavefront sensing and control. This input utilizes the same array structure, with the same frequency breakdowns, as the WFE instability. The values represent a WFE reduction factor, with 0 corresponding to complete WFE stabilization and 1 corresponding to no reduction of the WFE variance at all. The element-by-element product of the two (WFE instability and WFS\&C instability-transmission) arrays represent the final WFE instabilities that arrive at the coronagraph's focal-plane mask (FPM), $r_{ij}$.   
    \item \textit{Contrast sensitivity array ($s_{kj}$)}:  This is a 2D array of contrast sensitivity (change in contrast per change in RMS WFE) as a function of angular separation ($k$) and spatial mode ($j$) in ppt/pm.  
\end{enumerate}

We then combine these input arrays using the methodology described by Eq.~\ref{eq: deltacon} to get $\sigma_{\Delta C_k}$ and divide that by $\bar{C}_k$ to get the final value of $f_{\Delta C}$. This is broken out as a user-defined input parameter in EXOSIMS and so can be easily adjusted to perform ETC analyses. This process is depicted diagrammatically in Figure \ref{fig:flowchart}.

\subsection{The \texttt{ebs} Package}
\label{sec: ebs_package}
EBS is hosted on GitHub and is implemented as the Python 3 package \texttt{ebs} which includes a conda environment definition file \texttt{environment.yml}. It also includes the sample input YAML file \texttt{parameters.yml} which contains all of the input parameters EBS needs for both its computations and to initialize the EXOSIMS ETC.

EBS requires three 2D input arrays: WFEs, WFS\&C instability transmission factors, and sensitivities as described in \S \ref{sec: ebs} which are in the format of comma separated value (CSV) files. WFE and WFS\&C instability-transmission factors have rows corresponding to each temporal mode ($i$) and columns corresponding to each spatial mode ($j$) as described in Tables \ref{table:temp_mode_def} and \ref{table:spatial_mode_def}. For the sensitivities, rows correspond to the angular coordinate ($k$) and columns correspond to spatial mode ($j$). In all cases, the definitions of the temporal and spatial frequency bins come from those defined by the ULTRA Program\cite{2019CoyleULTRA}. Example CSV files are included in the \texttt{inputs} folder of the repository to provide the user with all of the requirements to run the base functionality of the code. 

The main functionality of \texttt{ebs} is contained in the \texttt{ErrorBudget} class which syntheses the input WFE, WFS\&C and sensitivity arrays into $f_{\Delta C}$ using the methods outlined in Section \ref{sec:core_func}, gets the EXOSIMS information from the \texttt{parameters.yml}, and calls the EXOSIMS ETC to output an exposure time. For more detailed implementation information, users should consult the \texttt{README.md} on the GitHub repository.  We also note that EBS's exposure-time calculation has been benchmarked against other ETCs \cite{2025ETCCrossCal}.

\begin{deluxetable}{lll}
\tablewidth{0pt}
\tabletypesize{\scriptsize}
\tablecaption{Temporal Mode Definitions \label{table:temp_mode_def}}
\tablehead{
\colhead{Name} & \colhead{Frequency (Hz)} & \colhead{Description} \\
}
\startdata
 Static  & 0 &  Static modes that do not change with time. \\
LF1  & $5.8\times 10^{-6}$ (48 hours) - 0.001 & Observing scenario and coronagraph high-order wavefront sensor bandwidth, depending on target-star brightness.  \\
LF2  & 0.001 - 0.01 & Coronagraph low order wavefront sensor (LOWFS) bandwidth, depending on target-star brightness. \\
LF3 & 0.01 - 1 &  Telescope alignment (Primary Mirror (PM)/Secondary Mirror (SM) rigid body motion). \\
MF & 1 - 10 &  PM segment-level, rigid-body sensing \& control. \\
HF & \textgreater 10 &  Uncontrolled modes or effects removed with image processing. \\
\enddata
\end{deluxetable}

\begin{deluxetable}{lll}
\tablewidth{0pt}
\tabletypesize{\scriptsize}
\tablecaption{Spatial Mode Definitions \label{table:spatial_mode_def}}
\tablehead{
\colhead{Name} & \colhead{Frequency (cycles/diameter)} & \colhead{Description} \\
}
\startdata
Low  & $\sim$ 2-4 (Zernike Noll 2-11) &  Global alignment of PM and low order PM modes (backplane). Can be compensated w/ SM motion \\
Mid  & 4-15 &  PM stabilizing assembly (PMSA) rigid body motion; low order PMSA modes.  \\
High  & 15-16 & PMSA mid spatial modes (i.e. mount print through). \\
High+ & \textgreater 60 &  PMSA high spatial modes above the DM correction range \\
\enddata
\end{deluxetable}

\subsubsection{Single Value Parameter Sweeps}
EBS contains the class \texttt{ParameterSweep} which is designed to provide an easy interface for performing single parameter studies to identify trends by updating an input of choice and iteratively running the ETC (see Section \ref{sec: ebs_example} for specific examples). This is further facilitated by the command line interface \texttt{run\_ebs} which allows users to run these single parameter sweeps from the terminal.  

\subsubsection{Multivariate Markov Chain Monte Carlo and Nested Sampling Sweeps}
\label{sec: mcmc_intro}

EBS also contains Markov-chain Monte Carlo and Nested Sampling modes which were created to survey required integration times in high dimensional spaces of observatory-coronagraph states (OCS). Mathematically, an OCS is a vector of values specifying observational and coronagraph-instrument characteristics that determine $t_{req}$ (e.g. spectral bandwidth, throughput and contrast as a function of angular separation, exo-zodi level, as well as the parameters listed in Tables \ref{table:mission_params}, $\delta_{ij}$, $s_{kj}$, and $\gamma_{ij}$, etc.). Our survey algorithm utilizes a MCMC or Nested Sampling approach, as implemented by either the \texttt{emcee}\cite{2013emcee} or \texttt{dynesty}\cite{dynesty_paper, dynesty_zenodo} packages respectively, to sample coronagraph states. 

A Markov chain has the property that, after a ``burn in" period, the chain generates values representative of those drawn from an intended (or ``posterior") probability distribution; for multi-variable PDFs, the Markov chain generates samples of state vectors (as opposed to scalar values). In addition, MCMC is innately independent of the initial state, and therefore, one obtains the same final distribution regardless of initial input values. In general, MCMC is more efficient in sampling large parameter spaces than uniform grid sampling because MCMC concentrates on sampling interesting regions in parameter space, as specified by the prior and likelihood functions. Moreover, the \texttt{emcee} method involves an ensemble of ``walkers," whose states evolve via an affine-invariant transformation algorithm \cite{2010goodman&weare} to achieve properties of MCMC sampling. The ensemble sampling enables parallel computing with approximately inverse-linear scaling of computing time with respect to the number of concurrent processes, same as that of embarrassingly parallelizable algorithms. The EBS MCMC mode ultimately produces a PDF (probability density function) that is the joint probability of input prior constraints on OCS variables and a ``likelihood" function based on $t_{req}$ (see below).       

Instead of estimating the posterior distribution, the Nested Sampling algorithm\cite{2004Skilling} instead primarily evaluates the Bayesian evidence as the the 1-D integral of the marginal likelihood integrated over the prior. To sample the parameter space, Nested Sampling takes a series of ``live points'', calculates their likelihoods, removes the lowest likelihood point, and then replaces it with a new live point, and continues. This preferential sampling of higher likelihood regions of parameter space makes Nested Sampling often much more efficient than MCMC. It is also less susceptible to getting stuck in local minima since it starts with live points drawn from a uniform distribution over the whole range of possible values. While detailed examples using the MCMC mode of EBS are provided in Section \ref{sec: mcmc_example}, we only provide a comparison study using the Nested Sampling mode in Section \ref{sec:nested_sampl} and leave more detailed explorations using and comparing the MCMC and Nested Sampling algorithms to future work. Both however are are readily available for users of EBS.

For both MCMC and Nested Sampling, we define our target PDF ($Pr$) as follows:  

\begin{equation}  
\label{eq: bayes}
Pr(\mathbf{x}|\mathbf{t}) \equiv L(\mathbf{t}|\mathbf{x})p(\mathbf{x})
\end{equation}

Here, $\mathbf{x}$ is a vector of user-selected OCS variables; see EXOSIMS documentation\cite{opticalsystem} for the full list of possible variables. The vector $\mathbf{t}$ comprises target integration-time values as a function of planet-star angular separation. The equation's right-hand side contains two PDFs; $p(\mathbf{x})$ is the prior probability of coronagraph-state variables, and $L(\mathbf{t}|\mathbf{x})$ is the likelihood of $\mathbf{t}$ conditioned on $\mathbf{x}$.  The prior function,  $p(\mathbf{x})$, specifies the distribution of parameter values to be explored, thus defining the extents of the OCS space to be surveyed, in a probabilistic manner.  The likelihood function, $L(\mathbf{t}|\mathbf{x})$, specifies the distribution of required-integration-time values to be of interest (i.e. the target range of acceptable integration times). 

For the MCMC, after burn-in, the Markov chain process produces a population of OCSs that represent the joint probability of the likelihood and prior functions: i.e. observatory-coronagraph states within the design space that yield required integrations times of interest.  The software stores the initial input parameters as well as the OCS vectors, $t_{req}$ and other key computed values associated with every state explored in a MCMC run. Example trade studies using this mode are outlined in Section \ref{sec: mcmc_example}.

\section{Sample ETC Studies with USORT Observatory Parameters}
\label{sec: ebs_example}

In this section we will go through a series of example studies that demonstrate the range of functionalities of EBS and provide insights on how certain observatory and instrument parameters affect exposure times for exo-Earth detection and characterization. For these studies, we will be using the observatory parameters defined by the Ultrastable Observatory Roadmap Team (USORT)\cite{USORT}. For the coronagraph we will be using an optical vortex coronagraph (OVC) designed for the USORT off-axis aperture with throughput curves calculated by the Coronagraph Design Survey\cite{2024CDS}. For the coronagraphic raw contrast, a fixed value of $10^{-10}$ was used at all angular separations unless otherwise specified. Exo-zodiacal light  was assumed to be constant as a function of angular separation. These and other parameters of interest are listed in Table \ref{table:mission_params}. 


\begin{deluxetable}{ccl}
\tablewidth{0pt}
\tabletypesize{\scriptsize}
\tablecaption{Coronagraph-based Mission Parameters\label{table:mission_params}}
\tablehead{
\colhead{Parameter} & \colhead{Value} & \colhead{Description} \\
}
\startdata
& & \bf{General Parameters} \\

$D$ & $7.87$ m & Telescope circumscribed diameter\\
$D_{\rm ins}$ & $6.5$ m & Telescope inscribed diameter\\
$A$ & $427518$ cm$^2$ & Collecting area of telescope\\
$X$ & $0.7$ & Photometric aperture radius in $\lambda/D_{\rm ins}$ \\
$\Omega$ & $\pi(X \lambda/D_{\rm ins})^2$ radians & Solid angle subtended by photometric aperture \\
\hline
& & \bf{Detection Parameters (\S 4.1)} \\
$\lambda_{\rm d}$ & $0.5$ $\mu$m & Central wavelength for detection \\
$\Delta\lambda_{\rm d}$ & $20\%$ & Bandwidth assumed for detection \\
S/N$_{\rm d}$ & $5$ & S/N required for detection \\
$T_{\rm optical,d}$ & $0.53$ & End-to-end reflectivity/transmissivity at $\lambda_{\rm d}$ \\
$\theta_{\rm pix,d}$ & 9.28 mas & scale of detector pixel for detections \\
\hline
& & \bf{Characterization Parameters (\S 4.2)} \\
$\lambda_{\rm c}$ & $1.0$ $\mu$m & Central wavelength for characterization \\
$\Delta\lambda_{\rm c}$ & $20\%$ & Bandwidth assumed for characterization \\
S/N$_{\rm c}$ & $5$ & S/N required for characterization (per spectral channel) \\
$l_{samp}$ & 2 & Number of pixels per lenslet row \\
$T_{\rm optical,c}$ & $0.3$ & End-to-end reflectivity/transmissivity at $\lambda_{\rm c}$ \\
$\theta_{\rm pix,c}$ & 9.28 mas & scale of detector pixel for detections \\
\hline
& & \bf{Detector Parameters (\S 4.1)} \\
$\xi$ & $3\times10^{-5}$ $\textrm{counts}$pix$^{-1}$ s$^{-1}$ & Dark current\\
RN & 0 $\textrm{counts}$ pix$^{-1}$ read$^{-1}$ & Read noise\\
$\tau_{\rm read}$& 1000 s & Time between reads\\
CIC & $1.3\times10^{-3}$ $\textrm{counts}$ pix$^{-1}$ frame$^{-1}$ & Clock induced charge\\
$T_{\rm QE}$ & $0.675$ & QE of the detector at all wavelengths \\
\hline
\enddata
\end{deluxetable}

\begin{deluxetable}{cccccc}
\tablewidth{0pt}
\tabletypesize{\scriptsize}
\tablecaption{Target Star and Planet Properties \label{table:stellar_params}}
\tablehead{
\colhead{HIP Number}  & \colhead{Spectral Type} & \colhead{dist (pc)} & \colhead{$M_v$} & \colhead{EEID (mas)} & \colhead{EEPSR} 
}
\startdata
32439 A  & F8V & 18.2  & 5.4 & 74.23 & 6.34e-11 \\
77052 A  & G5V & 14.8  & 5.9 & 61.74 & 1.39e-10 \\
79672  & G2Va & 14.1  & 5.5 & 73.99 & 1.06e-10 \\
26779  & K1V & 12.3  & 6.2 & 56.33 & 2.42e-10 \\
113283  & K4Ve & 7.6  & 6.4 & 58.29 & 5.89e-10 \\
\hline 
\enddata
\end{deluxetable}

Five fiducial stars spanning F, G, and K spectral types were chosen as representative target stars. They are all featured in both the HWO Preliminary Input Catalog (HPIC)\cite{2024HPIC} as well as the ExEP Mission Star List for the Habitable Worlds Observatory\cite{2024ExEPMissionStars} and key properties are summarized in Table \ref{table:stellar_params}. Planets were all assumed to be Earth-like with separations scaled by the host star luminosity. Other astrophysical assumptions are summarized in Table \ref{table:astroassumptions}. 

For these analyses we are using the sensitivities, WFE values, and WFS\&C factors shown in Tables \ref{table:wfe_wfsc_modes} and \ref{table:sensititivities_closing}. These values were selected based on analyses by ULTRA\cite{2019CoyleULTRA}, but scaled to yield reasonable exposures. Similar to the observatory parameters, these numbers will be refined as the mission develops and currently represent a best starting point to be able to perform parameter studies. With regards to the WFS\&C values, we would like to remind the reader that ``0'' indicates total suppression of the mode and ``1'' indicates no control. It can then be seen that despite the high values of temporally static WFE, this mode will have no impact at the coronagraph's focal plane mask due to the effect of the WFS\&C system.

It is important to note that for the following explorations the variables of interest are assumed to be independent. In reality, many of these coronagraph instrument components, such as throughput and post-processing efficiency, will be correlated. These relationships are beyond the scope of this work but are an interesting avenue of future exploration.

\begin{deluxetable}{ccl}
\tablewidth{0pt}
\tabletypesize{\scriptsize}
\tablecaption{Astrophysical Parameters\label{table:astroassumptions}}
\tablehead{
\colhead{Parameter} & \colhead{Value} & \colhead{Description} \\
}
\startdata
$R_{\rm p}$ & $1.0$ $R_{\rm Earth}$ & Planet radius \\
$a$ & $1.0$ AU & Planet semi-major axis\tablenotemark{a} \\
$e$ & $0$ & Orbital eccentricity \\
$\theta$ & $\pi/2$ & Phase angle \\
$\Phi$ & Lambertian & Phase function \\ 
$A_G$ & $0.2$ & Planetary geometric albedo\\
$z$ & 23 mag arcsec$^{-2}$ & V band surface brightness of zodiacal light \tablenotemark{b} \\
$z'$ & 22 mag arcsec$^{-2}$  & V band surface brightness of 1 zodi of exo-zodiacal dust \\
$n$ & $3.0$ & Exo-zodi level of each star \\
Habitable zone (HZ) & $(0.95 - 1.67) \cdot EEID$\tablenotemark{c} & Definition of the habitable zone\cite{1993Kasting1993} \\
\hline
\enddata
\vspace{-0.1in}
\tablenotetext{a}{For a solar twin.  This distance is scaled by $\sqrt{L_{\star}/L_{\odot}}$.}
\tablenotetext{b}{EBS assumes a fixed zodiacal light surface brightness regardless of pointing.}
\tablenotetext{c}{EEID: Earth-equivalent-insolation distance.}
\end{deluxetable}

\subsection{Raw Contrast and Wavefront Error}
\label{sec: wfe_contrast} 
\begin{figure}
    \centering
    \includegraphics[width=0.7\linewidth]{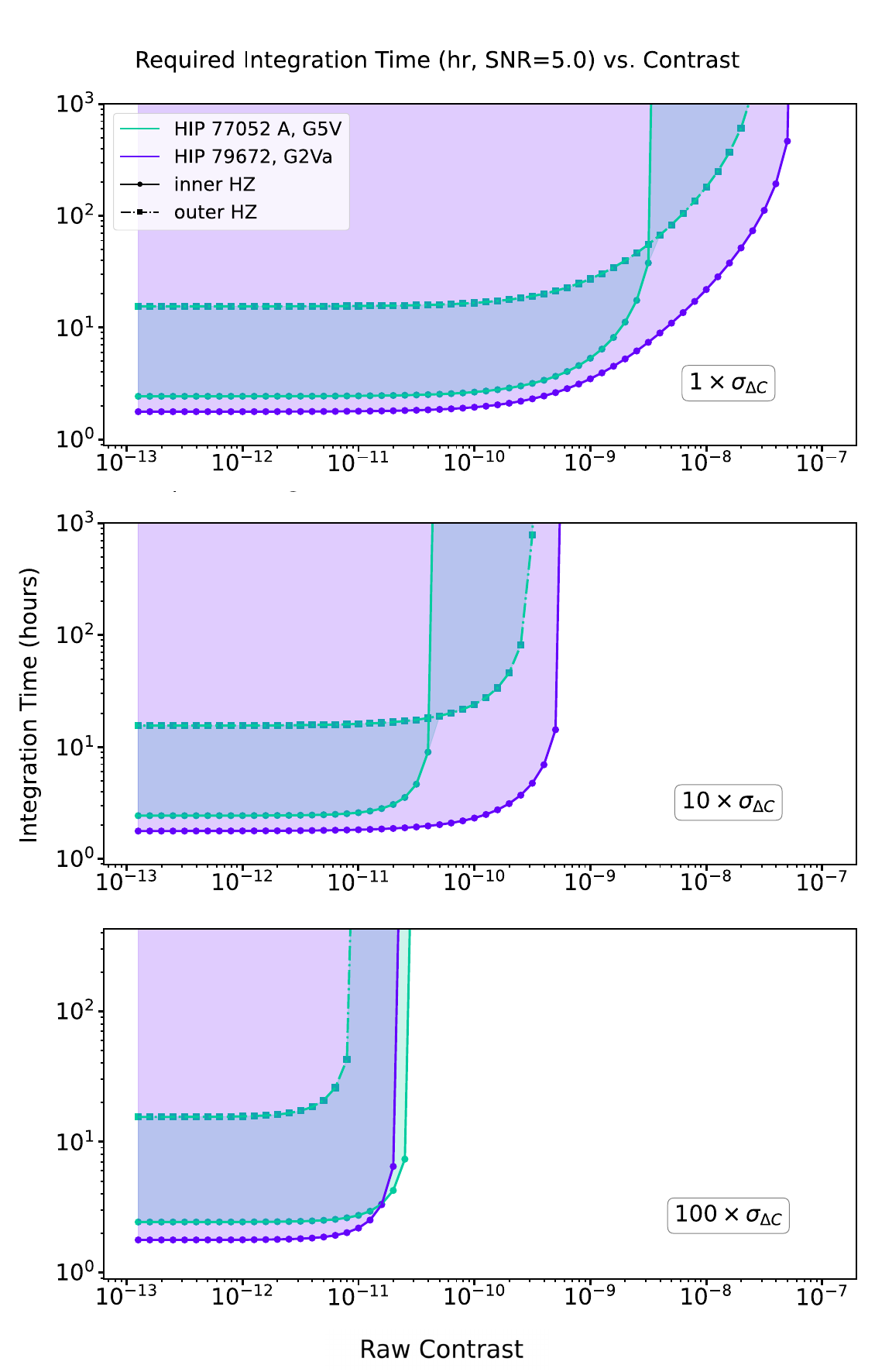}
    \caption{Required integration time for the detection of an Earth-like planet around two G-stars as a function of coronagraphic raw contrast. In each panel the two stars are represented by different colors with both the inner and outer habitable zones (HZ) displayed. The area between the inner and outer HZs is shaded. From top to bottom, the three panels represent increasing amounts of assumed WFE. The breakpoints where exposure times shoot to infinity indicate raw contrasts for which there is no possibility of detection, regardless of exposure time.}. 
    \label{fig:wfe_contrast}
\end{figure}

The relationship between coronagraph raw contrast and contrast stability is incredibly pertinent for HWO as both high-performance coronagraphs and ultra-stable observatory architectures are critical paths of ongoing technology development\cite{2024HWOFeinberg}. The break point defined in Eq.~\ref{eq:condition} is highly sensitive to introduced WFE instabilities as they increase the speckle residual noise which in turn increases $r_{\Delta I}$. This means that for a given raw contrast and WFS\&C instability-transmission factor, changing the input WFE shifts the allowed regions of parameter space where Earth-like planets are detectable. Decreasing the WFE (increasing telescope stability) relaxes the requirement on raw contrast and vice versa. This can be seen in Figure \ref{fig:wfe_contrast} where the required exposure times for exo-Earth detection are plotted as a function of the coronagraph raw contrast for a series of differing WFE instabilities \textemdash parameterized by applying a multiplicative factor ($n$) on $\sigma_{\Delta  C}$ (Eq.~\ref{eq: deltacon}). Here it can be seen that for larger introduced differential WFEs , the breakpoints where exposure times shoot to infinity are at deeper contrasts than for smaller WF drifts. It is important to note, as per Eq.~\ref{eq:sigma_delta_I}, this interplay is also influenced by more effective post processing, as well as through the development of coronagraphs that have less sensitivity to the introduced WFE at the temporal and spatial modes of interest. Both of these parameters were fixed for this exploration, but are expected to be objects of interesting future study. This illustrates an important feature of the interplay between raw contrast and wavefront drift. For fixed wavefront drifts, increasing the raw contrast increases exposure time in a continuous manner. For a fixed raw contrast, on the other hand, degrading wavefront stability can result in making some observations simply impossible, even with infinite exposure times, due to the singularity in Equation \ref{eq: ttsnr}.

\begin{deluxetable}{c|cccc}
\tablewidth{0pt}
\tabletypesize{\scriptsize}
\tablecaption{WFE and WFS\&C Spatio-temporal Mode Values \label{table:wfe_wfsc_modes}}
\tablehead{& \multicolumn{4}{c}{Spatial Modes} \\ 
Temporal Modes  & \colhead{Low} & \colhead{Mid} & \colhead{High} & \colhead{High+}
}
\startdata
& \bf{WFE [pm]} & & & \\ 
Static  & 20000 &  15000 &  5000 &  5000  \\
LF1  & 87.3 &  26.2 &  0.4 &  0.0  \\
LF2  & 87.3 &  26.2 &  0.4 &  0.0 \\
LF3 & 87.3 &  5.2 &  3.5 &  0.0\\
MF & 17.5 &  5.2 & 3.5 &  0.0  \\
HF & 17.5 &  5.2 &  3.5 &  0.0\\
\hline 
& \bf{WFS\&C} & & & \\ 

Static  & 0.0 & 0.0 & 0.0 & 0.0  \\
LF1  & 6.5E-3 & 9.1E-3 & 1.0 & 1.0 \\
LF2  & 7.3E-3 & 7.8E-3 & 1.0 & 1.0 \\
LF3 & 8.4E-3 & 1.3E-2 & 2.86E-1 & 1.0 \\
MF & 1.3E-2 & 9.4E-3 & 2.86E-1 & 1.0  \\
HF & 1.2E-2 & 1.5E-2 & 2.86E-1 & 1.0 \\
\hline
\enddata
\end{deluxetable}

\begin{deluxetable}{cc|cccc}
\tablewidth{0pt}
\tabletypesize{\scriptsize}
\tablecaption{Sensitivity Values [ppt/pm] \label{table:sensititivities_closing}}
\tablehead{\multicolumn{1}{c}{}&\multicolumn{1}{c}{}&\multicolumn{4}{c}{Spatial Modes} \\
\colhead{r (as)} & \colhead{r ($\lambda/D$)} & \colhead{Low} & \colhead{Mid} & \colhead{High} & \colhead{High+}
}
\startdata
0.035 & 2 & 6.4 &  10.0 &  0.0 &  0.0   \\
0.052 & 3 & 6.4 &  5.0 &  0.0 &  0.0   \\
0.070 & 4 & 0.0 &  3.0 &  0.0 &  0.0 \\
0.087 & 5 & 0.0 &  3.0 &  0.0 &  0.0 \\
0.104 & 6 & 0.0 &  1.5 & 0.0 &  0.0  \\
0.122 & 7 & 0.0 &  1.5 &  0.0 &  0.0 \\
\hline
\enddata
\end{deluxetable}

\begin{figure}
    \centering
    \includegraphics[width=0.7\linewidth]{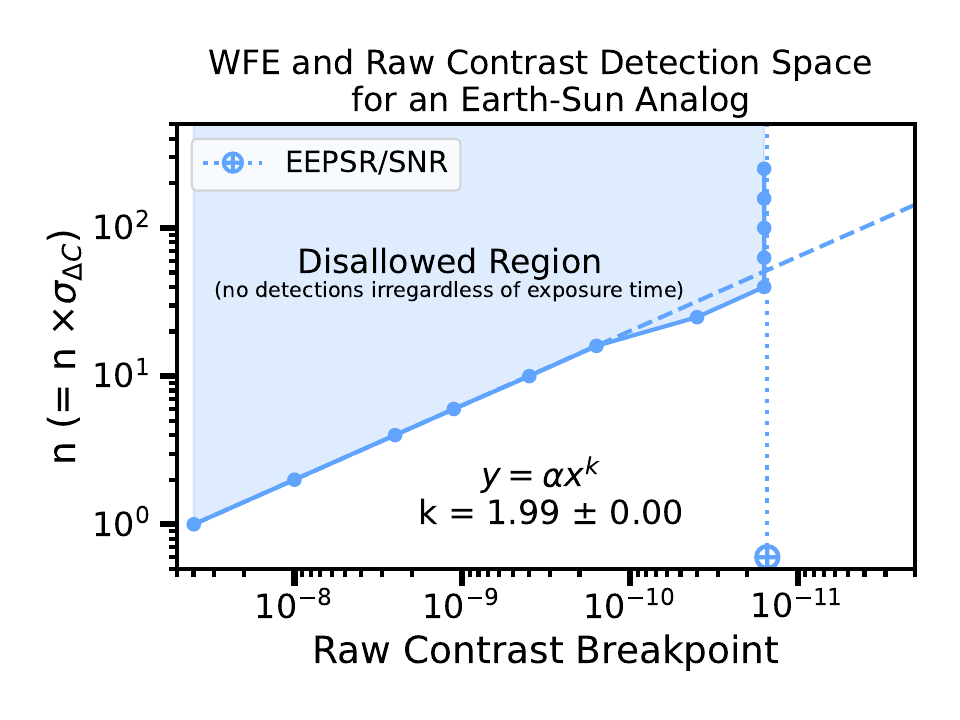}
    \caption{Contrast at which the required exposure time for an $SNR=5$ exo-Earth detection goes to infinity (the raw contrast breakpoint) as a function of the multiplicative factor applied to the input WFE ($\sigma_{\Delta C}$). Here the solid line is the data from EBS and the dashed line is the best fit power law. The region to the left and above the solid line represents a regime of coronagraphic raw contrast and WFE where no Earth-like planet can be found around this star, regardless of exposure time. The crossed circle and dotted line represents the $EEPSR$ scaled by the SNR for this target. The breakpoints remain constant once the raw contrast is lower than the $EEPSR/SNR$ meaning that an Earth-like planet can be detected at the desired SNR in single exposures.}
    \label{fig:contrast_breaks_single}
\end{figure}

\begin{figure}
    \centering
    \includegraphics[width=0.7\linewidth]{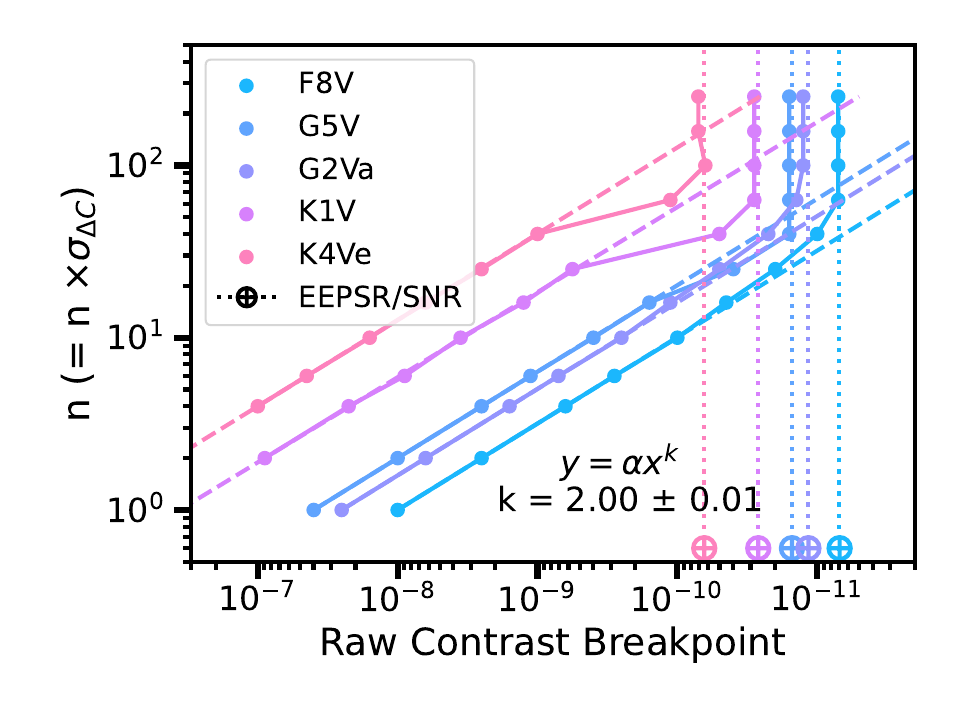}
    \caption{Same plot as Figure \ref{fig:contrast_breaks_single} but including data for all 5 fiducial target stars studied and removing the colored shaded region for clarity. As in the previous figure, regions to the left and above  these lines represent a regime of coronagraphic raw contrast and WFE where no Earth-like planet can be found around this star, regardless of exposure time.}
    \label{fig:contrast_breaks}
\end{figure}

Figure \ref{fig:contrast_breaks_single} shows the location of the breakpoints \textemdash contrasts at which no exposure times exist that can result in a planet detection \textemdash for the G5V star in Figure \ref{fig:wfe_contrast} as a function of $n \sigma_{\Delta C}$. This location should be tracked for each observatory trade considered because it signifies the ``point of no return" for a given architecture. This trend follows a power law with an exponent of 2 before reaching a contrast floor at $EEPSR/SNR$. This can be understood that once the raw contrast is deep enough that the planet can be acquired in single exposures with no subtraction required, then increasing the WFE instability no longer has any impact.
Figure \ref{fig:contrast_breaks} shows the same relationship for all 5 fiducial targets stars studied where the same trend is observed though the values of the individual curves are scaled depending on the host-star spectral type. The ``hook'' seen in the curves for planets around later spectral-type stars is likely due to the closer separations of these planets with respect to their host stars. Since the coronagraph sensitivities to wavefront errors are assumed to be much higher at smaller working angles (i.e. spatial frequencies, \emph{cf.} Table \ref{table:sensititivities_closing}), the required contrast needed to image these closer separation planets reaches an inflection point with increasing WFE before reaching the $EEPSR/SNR$ floor as indicated by the vertical dotted lines and crossed circles. 

\subsection{Spectral Resolution and Detector Noise}
\label{sec: det_noise_r}
\begin{figure}
    \centering
    \includegraphics[width=0.7\linewidth]{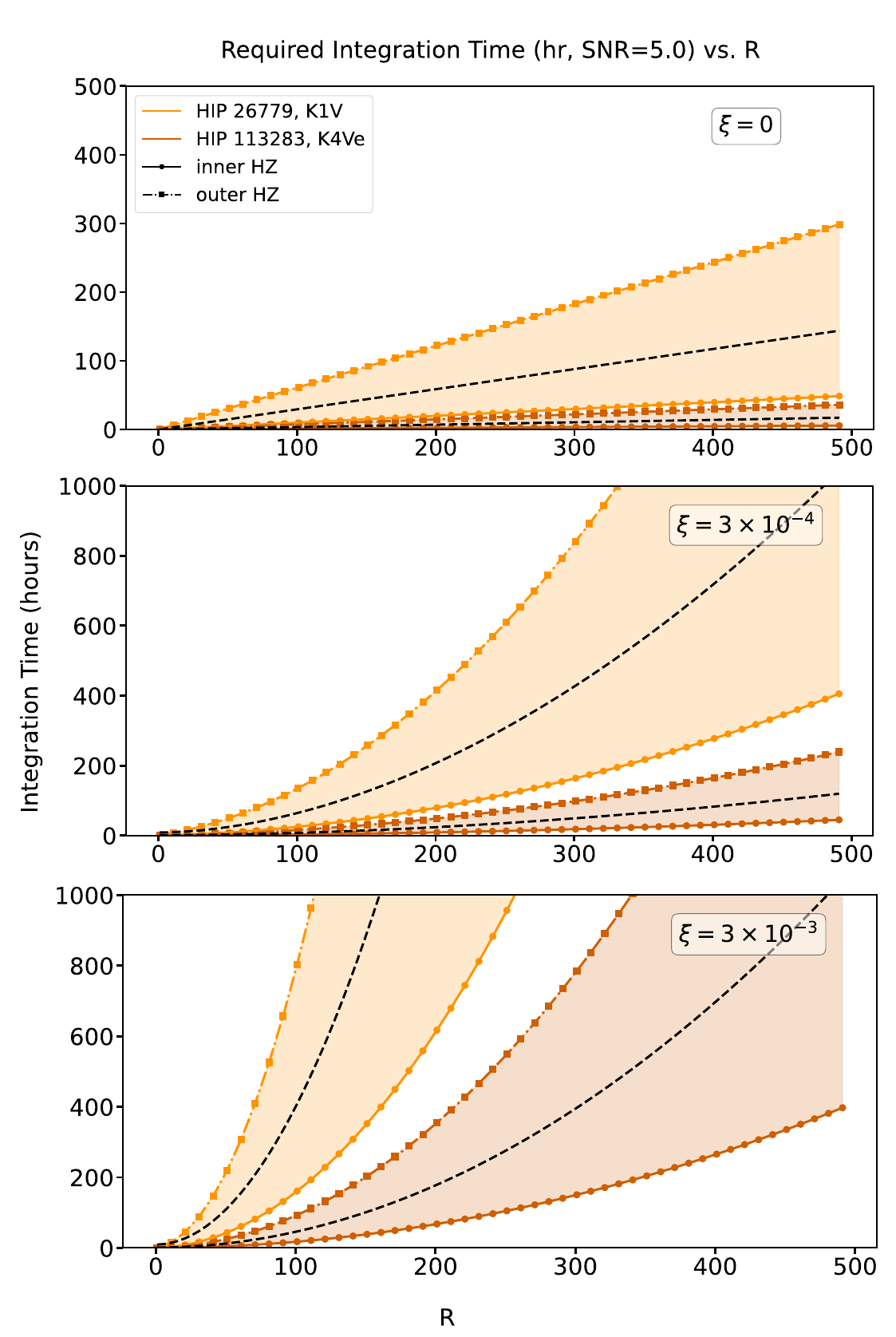}
    \caption{Sweeps of required integration time for $H_2O$ detection ($\lambda = 1000$ nm, SNR = 5) as a function of energy resolution ($R$) for a variety of values of dark current ($\xi$, counts/pix/s). Here all other detector noise terms are set to 0. As is also highlighted in Figure \ref{fig:det_noise_powers}, when dark current is increased, the relationship between $R$ and exposure time goes from linear to quadratic. The black dashed lines show the best fit power law following $y = \beta + \alpha x^k$.}
    \label{fig:det_noise_sweeps}
\end{figure}

In contrast to the systemic noise terms explored in the previous case study, detector noise is a source of random noise (contributing to $r_n$) and has no such break points. Though exposure times can increase rapidly and become practically unfeasible, there will always be an improvement to be gained through longer integrations. In this context, it is interesting to explore the case of spectroscopy where detector noise is more important than in the detection case since increasing the energy resolution ($R$) of an IFU spreads out light over a greater number of pixels, increasing the detector noise contribution. In this study, a lenslet sampling of 2 (4 lenslets per pixel) was used resulting in $\sim 25$ total pixels per spectral element, see also Table \ref{table:mission_params}. This is likely a pessimistic assumption for what will be the final design of the HWO IFU due to the following highlighted detector noise concerns. Figure \ref{fig:det_noise_sweeps} shows the relationship between integration time ($t_{req}$) and $R$ for $H_2O$ detection ($\lambda = 1000$ nm, SNR = 5) for three different values of dark current ($\xi$), setting other detector noise terms that rely on exposure time (read noise and clock-induced-charge) to 0. Here it can be seen that as dark current increases, the power law that best fits the relationship between $t_{req}$ vs. $R$ transitions from linear to quadratic. Note that these scaling laws are applicable only when considering SNR per spectral resolution element. A separate paper \cite{2025JB_MHRS}, uses EBS to identify the benefits of using a science metric that is based on the cross correlation with synthetic spectra of an exoplanet atmosphere, and another paper \cite{2026_Chen_ErrorBudget} presents an error-budgeting framework for coronagraphic spectroscopy.

\begin{figure}
    \centering
    \includegraphics[width=0.7\linewidth]{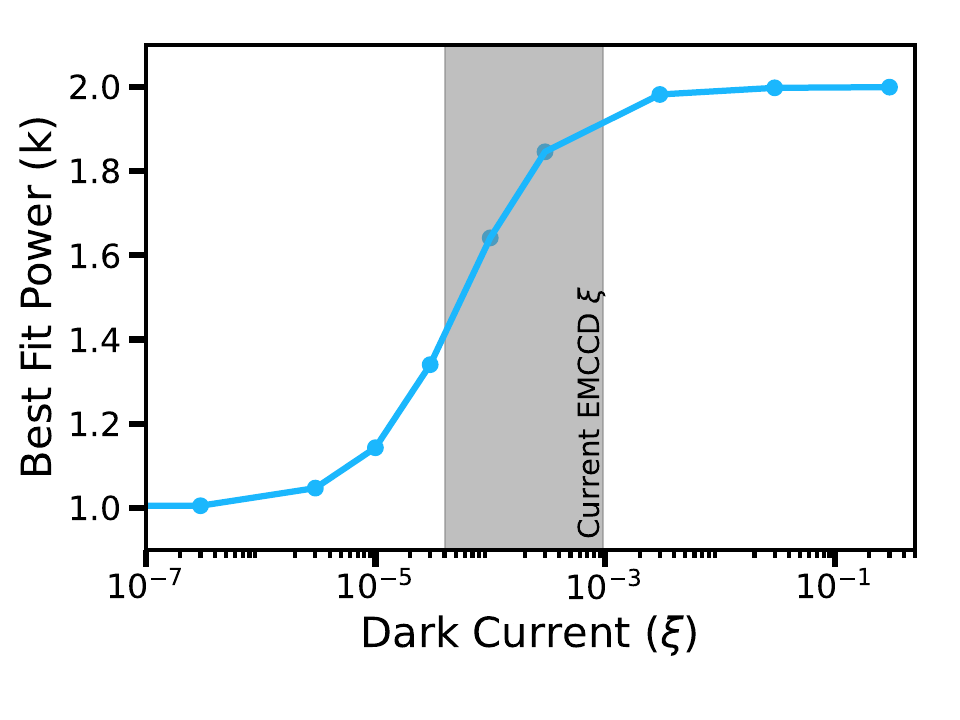}
    \caption{Exponent of the best fit power law ($y = \beta + \alpha x^k$) to the integration time as a function of energy resolution ($R$) at different values of dark current. For small values of dark current, the integration times increase linearly with $R$ ($k = 1$). At larger values of dark current, integration times increase quadratically with $R$ ($k = 2$) with a smooth transition between the two. The shaded region represents the approximate performance of current state-of-the-art EMCCD detectors to highlight that this transition occurs at a location where small improvements in detector noise properties can have a potentially large impact on exposure times for planet characterization. The lower bound of the shaded region is the dark current baselined by the LUVOIR\cite{2019LUVOIR} report and the upper bound is the expected performance for the Roman CGI detectors as found in the Roman CGI Noise Estimation Tool (\url{https://github.com/roman-corgi/cgi_noise})}
    \label{fig:det_noise_powers}
\end{figure}

These results are summarized in Figure \ref{fig:det_noise_powers} which shows the mean best fit power ($k$) for $t_{req}$ as a function of $R$ for different values of dark current ($\xi$). For low values of dark current, $t_{req}$ increases linearly with $R$ and at high values of dark current, $t_{req}$ increases quadratically with $R$ with a smooth transition between the two regimes. Interestingly, this transition occurs at dark current values that are close to current state-of-the-art EMCCD detector performance as is indicated by the shaded region.  

The reason for the transition between these two power laws can be understood by referencing back to Equation \ref{eq: ttsnr} and adapting it to apply to a single spectral bin. To do this we must include the energy resolution ($R$) which decreases astrophysical count rates by decreasing the effective bandpass per spectral channel. We also need to break the random noise ($r_n$) into two components. The first component includes noise counts due to astrophysical sources which will decrease with increased $R$: $r_{n, a}$. The second are noise contributions from the detector which will be independent of $R$: $r_{n, d}$. Making these substitutions, and assuming that the noise, planet, and residual speckles are all spectrally flat with respect to the resolution element, we get:

\begin{equation}
\label{eq: ttsnr_spec}
t_{spec} = \frac{\frac{r_{n, a}}{R} + r_{n, d}}{\left(\frac{r_{pl}}{SNR \cdot R}\right)^2 - \left({\frac{r_{\Delta I}}{R}}\right)^2} \, ,
\end{equation}

\noindent where we can expand the noise terms as

\begin{align}
\label{eq: noise_term_defs}
    r_{n, a} = r_{sr} + r_{z} + r_{ez} + r_{bl} \\
    r_{n, d} = r_{dc} + r_{rn} + r_{cic}
\end{align}

\noindent Here, $r_{sr}$ is the starlight residual count rate, $r_{z}$ is the zodiacal light count rate, $r_{ez}$ is the exo-zodiacal light count rate, $r_{bl}$ is the leaked binary light count rate, $r_{dc}$ is the dark current count rate, $r_{rn}$ is the read noise count rate, and $r_{cic}$ is the clock-induced-charge count rate. Here we assume Nyquist sampling at a certain design wavelength, so the number of pixels per spectral channel depends on wavelength, but not $R$. Plugging that into Eq.~\ref{eq: ttsnr_spec} and simplifying we get 

\begin{equation}
\label{eq: ttsnr_spec_full}
t_{spec} = \frac{SNR^2\left[R\left( r_{sr}+ r_{z} + r_{ez} + r_{bl} \right) + R^2\left( r_{dc} + r_{rn} + r_{cic}\right)\right] }{r_{pl}^2 - \left(SNR\cdot r_{\Delta I}\right)^2}
\end{equation}

Here it can be seen that if the sum of all of the detector noise terms are much less than astrophysical noise (i.e. $r_{dc} + r_{rn} + r_{cic} \ll r_{sr}+ r_{z} + r_{ez} + r_{bl}$), then the second term in the numerator is negligible and exposure times increase linearly. If detector noise dominates, then exposure times will increase like $R^2$ as is shown in Figure \ref{fig:det_noise_powers}.

\begin{figure}
    \centering
    \includegraphics[width=0.7\linewidth]{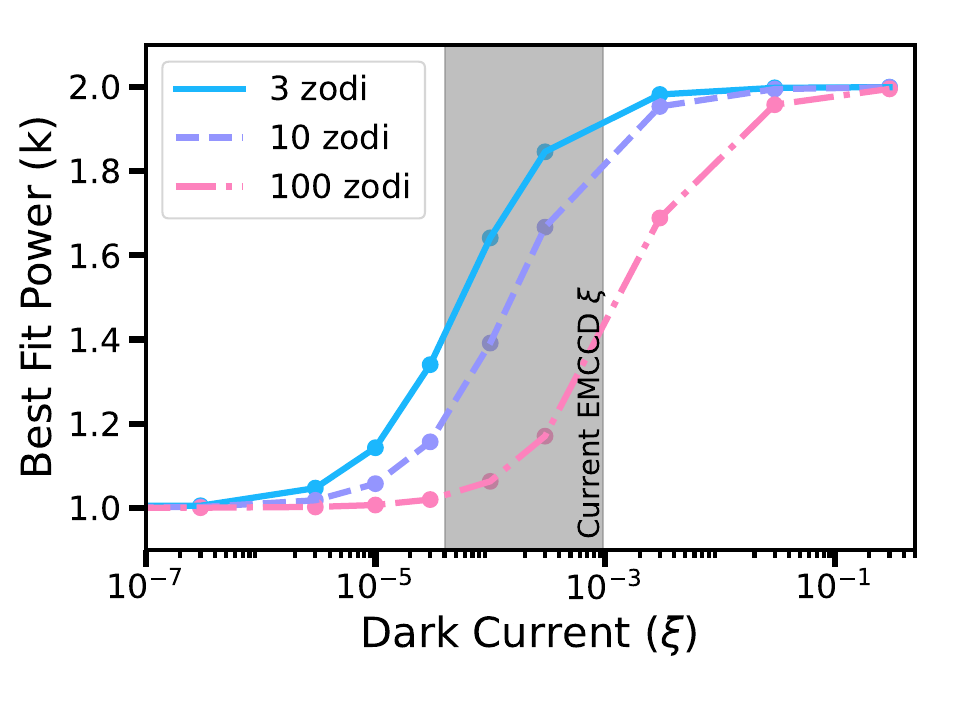}
    \caption{Exponent of the best fit power law ($y = \beta + \alpha x^k$) to the integration time as a function of energy resolution ($R$) as a function of dark current at varying exo-zodi levels. As the exo-zodi level is increased, the detector noise becomes less dominant and the curves can be seen to shift to the right.}
    \label{fig:det_noise_powers_zodi}
\end{figure}

\subsubsection{Varying exo-zodi levels}
\label{sec: exozodi}
One of the main sources of astrophysical noise for exo-Earth detection will be exo-zodiacal light. Analogous to the zodiacal light in our own Solar System, exo-zodi values around nearby stars can vary dramatically with values ranging from 0 to 2000 zodis (with a median value of 3 around Sun-like stars)\cite{2020HOSTS}. We performed the same sweep of exposure time as a function of $R$ at different values of dark current while also varying the exo-zodi brightness around the targets of interest. As can be seen in Figure \ref{fig:det_noise_powers_zodi}, as exo-zodi is increased, the astrophysical shot noise increases ($r_{n,a}$, Eq.~\ref{eq: noise_term_defs}). This in turn causes exposure times to increase linearly with $R$ for larger values of dark current as compared to the lower exo-zodi case as the exo-zodi shot noise will dominate the error budget ($r_{dc} + r_{rn} + r_{cic} < r_{sr}+ r_{z} + r_{ez} + r_{bl}$). Given our stated assumptions, from a spectral resolution standpoint there is not much benefit to improving detector noise from the current state-of-the-art for stars with greater than $\sim$ 100 exo-zodis.  
 
\subsection{High Dimensional Exploration using Markov-chain Monte Carlo and Nested Sampling}
\label{sec: mcmc_example}
Complementary to sweeps of individual parameters, EBS's MCMC and Nested Sampling modes enable high-dimensional exploration of OCS space to identify breakpoints in coronagraph performance and correlations between design parameters, as well as building databases of OCSs that meet integration-time criteria.  Moreover, such explorations can confirm the generality of conclusions drawn from single parameter sweeps (i.e. that the conclusions are not particular to the local, scalar nature of the sweep). Here we present two examples that are relevant to HWO assumptions, imaging/detection, and $R=70$ spectroscopy.    
\subsubsection{OCS Variables and Prior PDFs}
\label{sec: mcmc_priors}
For the imaging study, an MCMC run explored combinations of 23 OCS variables:  detector dark current, 10 WFS\&C instability-transmission values ($\gamma_{i,j}$), contrast in 6 angular-separation bands, and throughput in 6 angular-separation bands.  Tables \ref{table:prior_contrast_throughput_sens} and \ref{table:prior_wfe_wfsc} list the prior PDFs for these variables. In addition, we used a uniform prior PDF over the range (1.0E-6, 1.0E-4) counts/pixel/s for the camera dark current. Table \ref{table:prior_contrast_throughput_sens} shows that the run explored broad ranges of contrast and core-throughput. Table \ref{table:prior_wfe_wfsc} shows that we explored broad ranges of WFS\&C performance for low- and mid-spatial modes, while assuming fixed values for other modes. All other run inputs were as shown in Tables \ref{table:mission_params} - \ref{table:sensititivities_closing}. 

For the spectroscopy studies, we focused on only four key parameters: dark current, WFS\&C instability-transmission factor for the LF3-Mid temporal-spatial-frequency bin, and raw contrast and core throughput at 0.0785 arcsec working angle with the same priors as the imaging case.

\begin{deluxetable}{c|c|c|cccc}
\tablewidth{0pt}
\tabletypesize{\scriptsize}
\tablecaption{Prior Contrast \& Core Throughput PDFs \label{table:prior_contrast_throughput_sens}}
\tablehead{Angular Separation [arcsec] & Contrast & Core Throughput & \multicolumn{4}{c}{Sensitivity [ppt/pm]} \\ 
& & & \colhead{Low} & \colhead{Mid} & \colhead{High} & \colhead{High+}}
\startdata
0.0435  & (1E-11, 1.0E-9) &  (0.01, 0.6) & 6.4 & 10 & 0 & 0  \\
0.061  & (1E-11, 1.0E-9) &  (0.01, 0.6) & 6.4 & 5 & 0 & 0  \\
0.0785  & (1E-11, 1.0E-9) &  (0.01, 0.6) & 0 & 3 & 0 & 0 \\
0.0955 & (1E-11, 1.0E-9) &  (0.01, 0.6) & 0 & 3 & 0 & 0 \\
0.113 & (1E-11, 1.0E-9) &  (0.01, 0.6) & 0 & 1.5 & 0 & 0  \\
0.1305 & (1E-11, 1.0E-9) &  (0.01, 0.6) & 0 & 1.5 & 0 & 0 \\
\hline 
\enddata
\vspace{-0.8cm}
\tablecomments{Parentheses indicate uniform distributions over (exclusive) ranges used as prior PDFs for the MCMC run.}  
\end{deluxetable}

\begin{deluxetable}{c|cccc}
\tablewidth{0pt}
\tabletypesize{\scriptsize}
\tablecaption{Prior WFS\&C Instability-Transmission Values \& PDFs \label{table:prior_wfe_wfsc}}
\tablehead{& \multicolumn{4}{c}{Spatial Modes} \\ 
Temporal Modes  & \colhead{Low} & \colhead{Mid} & \colhead{High} & \colhead{High+}
}
\startdata
& \bf{WFS\&C} & & & \\ 
Static  & 0.0 &  0.0 &  0.0 &  0.0  \\
LF1  & (1E-6, 1) &  (1E-6, 1) &  1.0 &  1.0  \\
LF2  & (1E-6, 1) &  (1E-6, 1) &  1.0 &  1.0 \\
LF3 & (1E-6, 1) &  (1E-6, 1) &  0.286 &  1.0 \\
MF & (1E-6, 1) &  (1E-6, 1) & 0.286 &  1.0  \\
HF & (1E-6, 1) &  (1E-6, 1) &  0.286 &  1.0 \\
\hline
\enddata
\vspace{-0.8cm}
\tablecomments{Parentheses indicate uniform distributions over indicated ranges used as prior PDFs for the MCMC run.  Otherwise, numbers indicate fixed values.}
\end{deluxetable}

\subsubsection{Likelihood Function}
\label{sec:  mcmc_likelihood}
As discussed in Section \ref{sec: mcmc_intro}, likelihood is essentially a merit function of required integration time. For each of the $M$ fiducial stars, the MCMC and Nested Sampling algorithms compute a required integration time $t_n$ at each of $N$ specified angular separations (cf. left-most column in Table \ref{table:prior_contrast_throughput_sens}), and it also computes the mean required integration time:  
\begin{equation}
    t_{avg} = \frac{1}{MN}\sum_{m=1}^M{\sum_{n=1}^N{t_{mn}}}.
\end{equation}  
In the herein-presented example, $M=5$ and $N=6$. We then define the likelihood ($\mathcal{L}$) as follows:  
\begin{equation}  
\label{eq: likelihood}
\mathcal{L}(t_{avg}) = e^{-\frac{1}{2}\left(\frac{t_{avg}-t_0}{\sigma}\right)^2 }, \;
t_{avg} \ge 0
\end{equation}
This is a Gaussian distribution where $t_0$ and $\sigma$ define the center and the width of the distribution,  respectively. We enforce that the average integration time must not be a negative number.  In this example, $t_0 = 3.0$ days and $\sigma = 2.0$ days, and thus, the MCMC and Nested Sampling explorations concentrated on surveying observatory-coronagraph states that produce required integration times in the range of one to five days for the five fiducial stars.

\subsubsection{MCMC Imaging Results}
\label{sec: mcmc_results}
\begin{figure}
    \centering    
    \includegraphics[width=0.6\linewidth]{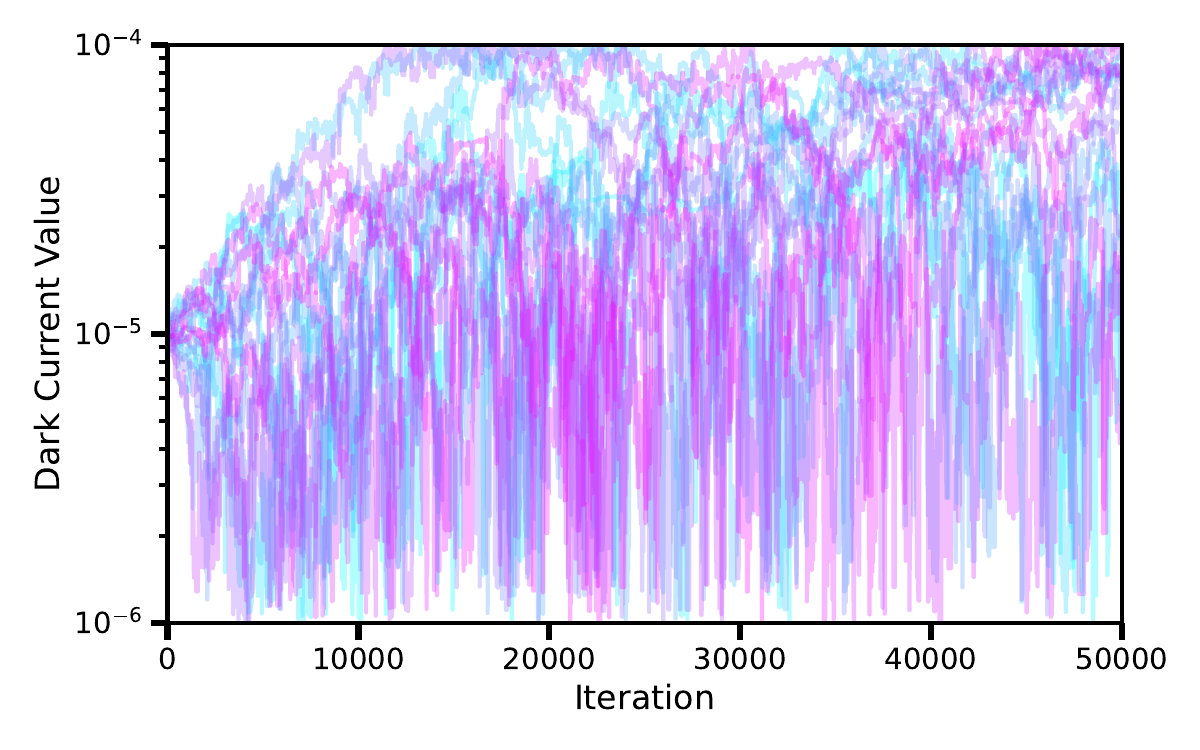}
    \caption{Evolution of the dark current value for 20 of the 100 total walkers vs. step number. Each colored trace represents a walker. After $\sim$ 14,000, steps the walkers appear to be fully burnt in from their median starting value of $1 \times 10^{-5}$.}
    \label{fig:tracks}
\end{figure}

This MCMC run employed 100 walkers and completed 50,000 steps. As such, the run explored 5 million OCSs. To ensure truly statistical sampling of the intended parameter space, the runs need to achieve burn-in. Autocorrelation length is one quantitative estimate of burn-in length \cite{2013emcee}.  Generally, the Markov chain must take a few times autocorrelation length (in the number of steps) to ensure all walkers have forgotten their initial states and explore the parameter space in a ``fully mixed'' fashion. We computed the autocorrelation length for each of the 23 OCS variables, with the median autocorrelation length being 2,478 steps, and the maximum 2,912.  Figure \ref{fig:tracks} plots the evolution tracks of the variable with the longest autocorrelation length, the dark current, associated with the 20 of the 100 walkers. After taking $\sim 1.4\times10^4$ steps, about five times the autocorrelation length, the walkers' ranges of motion have expanded fully, no longer influenced by the initial states.  This demonstrates that the run achieved fully statistical sampling of the parameter space as prescribed by the posterior PDF.  

\begin{figure}
    \centering
    \includegraphics[width=0.75\linewidth]{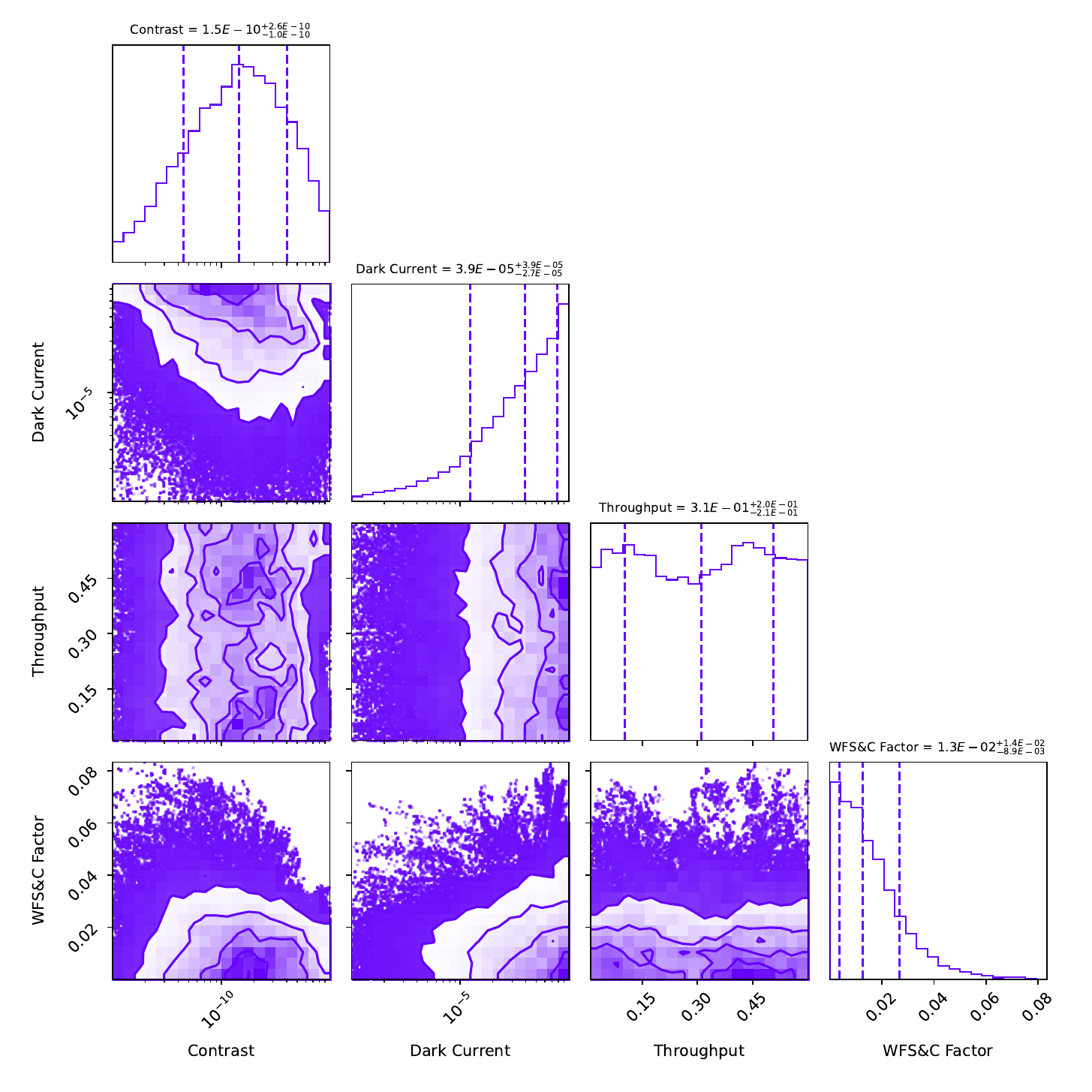}
    \caption{Corner plot for the exoplanet detection case of four select variables:  dark current, WFS\&C instability-transmission factor for the LF3-Mid temporal-spatial-frequency bin, raw contrast at 0.0785 arcsec working angle, and core throughput at the same working angle. The histograms of variables are shown along the ``hypotenuse" of the triangle panel pattern, and the other panels show joint probability densities between pairs of variables. The purple-dot density is proportional to the occurrence frequency of OCSs according to the posterior PDF. The dashed lines in the histograms and the contour lines in the 2-D joint plots mark population quartiles. The numbers above the histograms note the median$\pm$quartile variable values.}
    \label{fig:corner}
\end{figure}

Figure \ref{fig:corner} shows a corner plot of the resulting MCMC samples for four of the 23 variables:  dark current, WFS\&C instability-transmission factor for the LF3-Mid temporal-spatial-frequency bin, raw contrast at 0.0785 arcsec working angle, and core throughput at the same working angle. A corner plot shows the histogram of each variable (along the ``hypotenuse" of the triangle) and joint probability densities between pairs of variables. The purple-dot density is proportional to the occurrence frequency of OCSs according to the posterior PDF. The dashed lines in the histograms and the contour lines in the 2-D joint plots mark population quartiles. The numbers above the histograms note the median$\pm$quartile variable values.

Let us examine the contrast variable to elucidate the interpretation of these results. If one sets a design requirement that the raw contrast performance (at 0.0785 arcsec working angle) must be $\ge 1.5\times10^{-10}$, the median variable value, then only 50\% of the probable combinations (OCSs) of all other design variables (as dictated by the prior PDFs) can produce acceptable required integration times, as dictated by the likelihood function. In other words, the marginalized probability is 50\%. If one now sets this contrast requirement to be $\ge (1.5 + 3.0) \times 10^{-10}$, the upper quartile value, then the fraction of probable OCSs drops to 25\%. Naturally, the further one degrades the contrast performance, the smaller the fraction of probable OCSs, eventually to a point where no \emph{a priori} probable observatory-coronagraph design can produce required integration times that meet the likelihood condition.  We will refer to the 25-percentile variable value as the "quartile degradation point," or QDP. Table \ref{table: posterior} lists the median and QDP values for the 23 explored variables.  

Note that, going in the direction of improving contrast, the histogram peaks at around $1 \times 10^{-10}$ and then drops off sharply again toward smaller (better) contrast values. This is an artifact of the fact that we specified a Gaussian likelihood function centered at $t_0 = 3.0$ days; by design, the run effectively excluded sampling possible designs that produced very short ($\ll 3$ days) required integration times, although they would be highly desirable. Knowing that achieving coronagraphic performance for Earth-like exoplanets is technologically challenging, we imposed this condition to more densely sample the region near the upper limit of tolerable required integration times per mission scheduling. Therefore, our discussions below will focus on only the contours in the direction of longer required integration.  

We now make some more subtle observations to further elucidate interpretation of MCMC results.  
\begin{description}
    \item[WFS\&C instability-transmission factors] The histogram for the instability-transmission factor has a sharper drop-off than other parameters.  We attribute this to the fact that when $r_{\Delta I} \ge r_{pl}/SNR$, no physical solution exists, as shown in Equation \ref{eq: ttsnr}.  If we ran the MCMC exploration using only one target star at a single working angle, we would see a sharp break point beyond which no OCS solution exists. This confirms the findings of Section \ref{sec: wfe_contrast}, which were obtained via a single parameter sweep. When carrying out our MCMC based  global exploration of the trade space, we found that there is no combination of other mission parameters that can ``make up'' for the fundamental cliff associated with wavefront stability identified in Figure \ref{fig:wfe_contrast}.  
    \item[Dark Current and Core Throughput] Dark current and core throughput have relatively flat histograms, implying low sensitivity of required integration time to these variables. While this is true for the particular run setup, we note that core throughput is a critical factor in low photon-count scenarios such as spectroscopy (see Section \ref{sec:mcmc_spectro}), and sensitivity to detector noise can transition to different power laws depending on detector and astrophysical noise levels (\emph{cf.} Section \ref{sec: det_noise_r}). 
    \item[Design Trade between Contrast and WFS\&C]  The quartile contours of the joint-probability plot between contrast and WFS\&C instability-transmission factor has a generally negative $\Delta C/\Delta \gamma$ trend. We attribute this to the heterodyne term in Equation \ref{eq:ref_sub}.  For an incremental relaxation in WFS\&C-damping requirement one can restore the posterior probability by improving the raw contrast, and vice versa. The near vertical contour lines indicate that a relatively large change in contrast is required to compensate a change in wavefront stability, which makes sense because contrast scales quadratically with wavefront amplitude. 
    \item[Tolerable Residual WFE Instability at the FPM]  Using Equation \ref{eq: r_ij} with values in Tables \ref{table:wfe_wfsc_modes} and \ref{table: posterior}, we can compute the (post-WFS\&C residual) WFE instability incident on the FPM ($r_{ij})$.  Doing so and then root-square-summing all the Low and Mid spatial frequency bins yields a value of 3 pm. Moreover, Equation \ref{eq: deltacon} with sensitivity values in Table \ref{table:sensititivities_closing} allow us to compute the resulting contrast instabilities, which ranges from $\sim 6 \times 10^{-11}$ at the two smallest working angles to $\sim 5 \times 10^{-12}$ at larger working angles. Note that the implementation of post-processing will further relax the wavefront stability requirements. These wavefront and contrast stability values are in general agreement with HWO design goals.        
\end{description}

\begin{deluxetable}{c|c|c|ccc}
\tablewidth{0pt}
\tabletypesize{\scriptsize}
\tablecaption{Median \& QDP Values of Posterior PDFs \label{table: posterior}}
\tablehead{Angular Separation [arcsec] & Contrast $\times 10^{-10}$ & Core Throughput & \multicolumn{3}{c}{Instability Transmission Factor} \\ 
& & & & \colhead{Low} & \colhead{Mid}}
\startdata
0.0435  & 2.9 (5.3) & 0.29 (0.15) &  &  &  \\
0.061  & 2.3 (4.1) &  0.32 (0.17) & LF1 & 0.005 (0.009) & 0.012 (0.020) \\
0.0785  & 1.5 (3.0) & 0.31 (0.15) & LF2 & 0.006 (0.011) & 0.013 (0.022)  \\
0.0955 & 2.3 (4.3) & 0.26 (0.14) & LF3 & 0.006 (0.010) & 0.045 (0.076) \\
0.113 & 2.7 (4.7) &  0.26 (0.14) & MF & 0.019 (0.036) & 0.046 (0.078)  \\
0.1305 & 2.8 (5.2) &  0.27 (0.14) & HF & 0.026 (0.044) & 0.046 (0.075) \\
\hline 
\enddata
\vspace{-0.8cm}
\tablecomments{QDP values appear in parentheses.  The median and QDP values for dark current are $3.9 \times 10^{-5}$ $(1.9 \times 10^{-5})$}.
\end{deluxetable}

\subsubsection{MCMC Spectroscopy Results}
\label{sec:mcmc_spectro}
We also used the EBS MCMC mode to explore the spectroscopy case with an $R=70$ IFU using the same parameters as described in Section \ref{sec: mcmc_results}. For this study, to save on computation time, we only varied four select parameters of interest (dark current, WFS\&C instability-transmission factor for the LF3-Mid temporal-spatial-frequency bin, and raw contrast and core throughput at 0.0785 arcsec working angle) with 100 walkers and completed 10,000 steps. Since the average autocorrelation length for these 4 variables is 85 iterations, burn-in is achieved much quicker and we can safely use 1/5 the amount of iterations as the imaging study.

\begin{figure}
    \centering
    \includegraphics[width=0.75\linewidth]{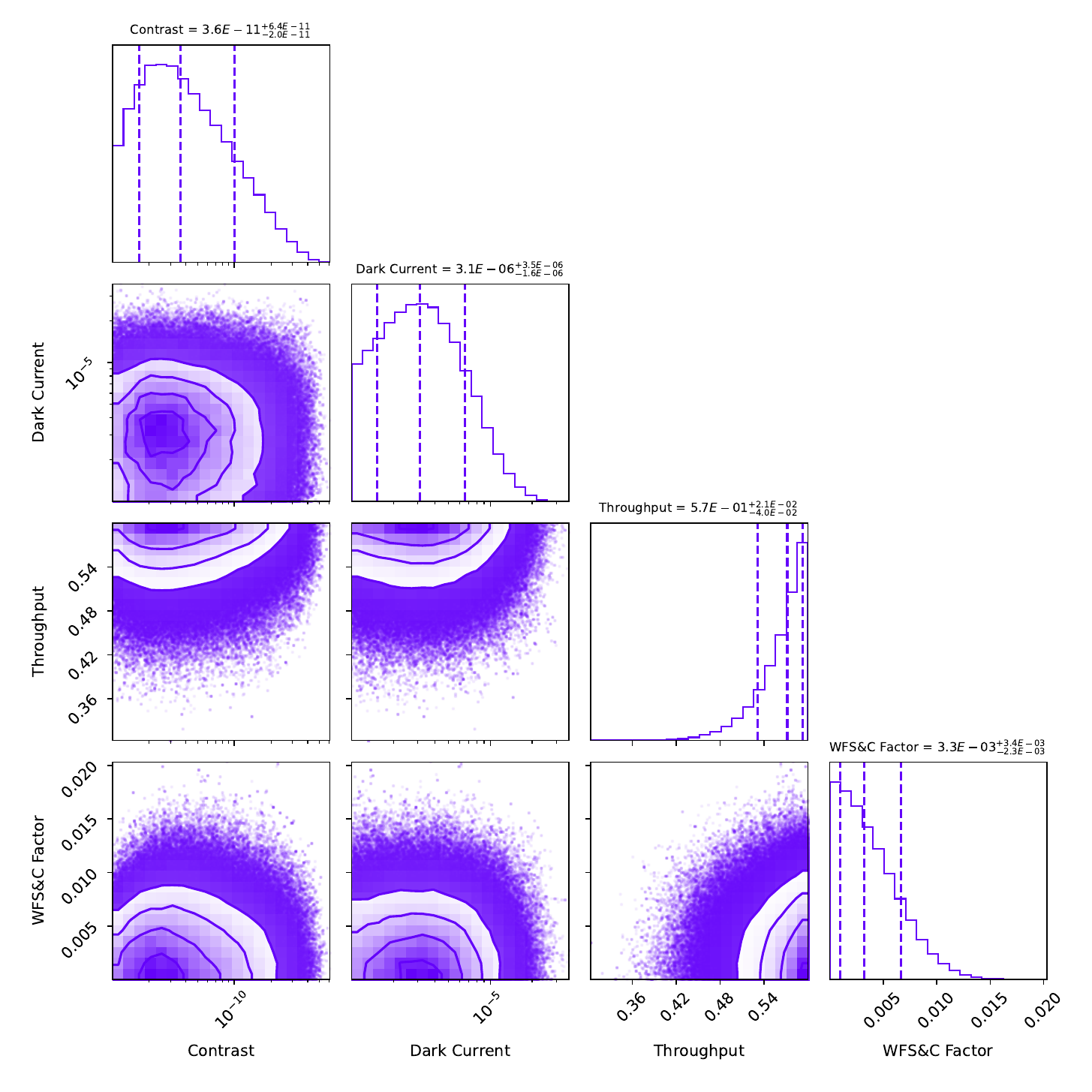}
    \caption{Corner plot using the same parameters as described for the imaging case in Section \ref{sec: mcmc_results} (Figure \ref{fig:corner}), but for an $R=70$ IFU.}
    \label{fig:corner_spectro}
\end{figure}

Figure \ref{fig:corner_spectro} shows our results in the same format as Figure \ref{fig:corner}. Comparing these two figures reveal key differences between the spectroscopy and imaging cases. First, all of the parameters of interest have much fewer allowable OCSs in the spectroscopy case further demonstrating that spectroscopy will drive the technology requirements for HWO. 
In particular, we can see that the median contrast decreases to $3.7 \times 10^{-11}$ which is approximately the value of the mean $EEPSR/SNR$ for the 5 targets of interest ($4.6 \times 10^{-11}$). This corroborates the results found in Section \ref{sec: wfe_contrast} where it was determined that there is no gain to be had in decreasing the raw contrast past this value. 
The other major difference between the imaging and spectroscopy cases is in the relationship between throughput and dark current. These values are each individually much more tightly constrained with median values for the dark current of $ 3.1\times 10^{-6}$ counts/pix/s and $5.7\times 10^{-1}$ for the throughput. Additionally, the correlation between throughput and dark current is clearly seen in the joint probability density where higher throughput values allow for higher values of dark current and vice versa.

\subsubsection{Nested Sampling Spectroscopy Comparison}
\label{sec:nested_sampl}

\begin{figure}
    \centering
    \includegraphics[width=0.75\linewidth]{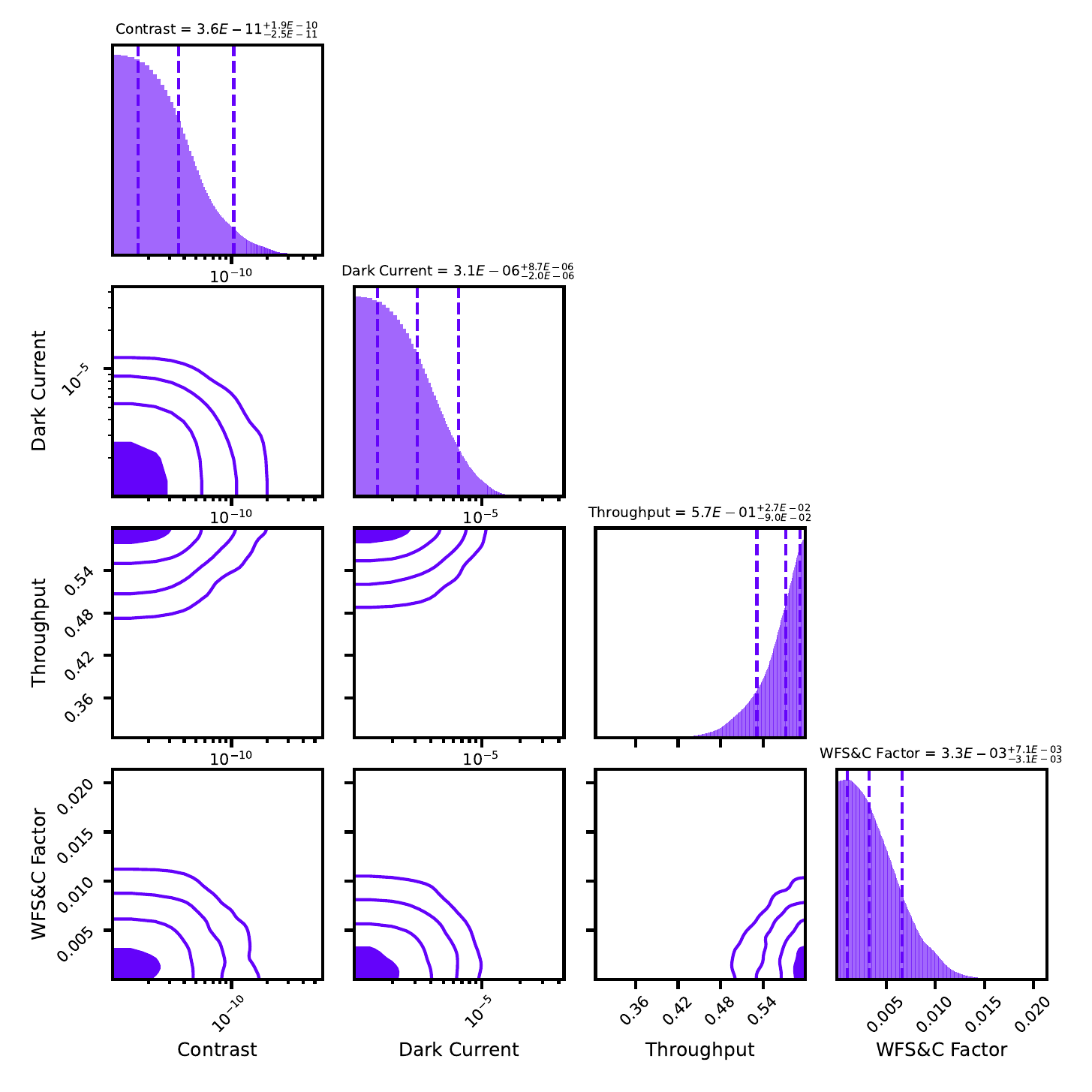}
    \caption{Corner plot results using identical parameters to Section \ref{sec:mcmc_spectro} (Figure \ref{fig:corner_spectro}) but using EBS's Nested Sampling functionality.}
    \label{fig:nested_sampling}
\end{figure}

To demonstrate EBS's Nested Sampling mode and cross-validate it with the MCMC method, we performed a multivariate sweep using identical parameters as the MCMC sweep in Section \ref{sec:mcmc_spectro} with results shown in Figure \ref{fig:nested_sampling}. Here it can be seen that we verify the MCMC spectroscopic results recovering the same median values for all parameters considered.

The above explorations demonstrate that the EBS's MCMC and Nested Sampling modes are useful for identifying \emph{design break points, sensitivities, and trades}, and the findings are compatible with those drawn from single-parameter sweeps as well as physical insights.



\section{Discussion}
\label{sec: Discussion}

EBS was validated against the EXOSIMS and AYO ETCs as part of an exposure time calculator cross-calibration effort\cite{2025ETCCrossCal} led by the Exoplanet Science Yields sub-Working Group under the HWO START/TAG before the formation of the HWO TMPO. As was the case for that work, for the parameter studies in Section \ref{sec: ebs_example} we selected 5 fiducial target stars as our emblematic sample to perform our calculations. In general, caution should be taken in drawing conclusions from small target samples as minor changes in stellar properties can have a potentially large impact on expected exposure. For this reason we have focused our results not on absolute exposure times, but instead on identifying general scaling laws and trends that are more robust and agnostic of specific target. For these analyses we also had to make assumptions about the observatory which are unlikely to represent the final design for an exo-Earth imaging mission such as HWO. As various coronagraph instrument and observatory trades are performed, this multivariate design space will be narrowed down with the aid of tools such as EXOSIMS and EBS. What's more, while we have focused on HWO exo-Earth imaging as the subject of this work, the EBS software and the mathematical formalisms described in Sections \ref{sec: assump_deriv} and \ref{sec: ebs} are agnostic to observatory or coronagraph instrument architecture and so can be applied more generally to any future coronagraph misson. 

When looking at the relationship between WFE and raw contrast in Section \ref{sec: wfe_contrast} we only explored changing the input WFE unilaterally across all spatial and temporal bins. In actuality, not all spatial and temporal frequencies are equally important and there will be some that will be harder to mitigate than others. Different coronagraph designs will also be sensitive to different types of these modes leading to a complex interplay between these parameters as was seen in the MCMC results in Section \ref{sec: mcmc_example}. Future work will involve further interrogation of these spatial and temporal frequencies individually to gain insight on what instabilities are of the most importance to mitigate in hardware or through active WFS\&C. 

With regards to the detector noise analysis in Section \ref{sec: det_noise_r}, in this work we have been operating under the assumption that the best metric for exo-Earth characterization is achieving a given SNR per Nyquist-sampled spatial and spectral bin. This is a conservative assumption due to the fact that, in very low count rate regimes (where detector noise is of most concern), it is not unreasonable to undersample in the spectral or spatial dimension as is the case for the JWST NIRSpec IFS \cite{2022Boker}. Additionally, for medium to high resolution spectrographs, ($R \sim 100-10,000$) where post-processing techniques such as template matching with cross correlation functions (CCF)\cite{2013Konopacky, 2019JB} can be used, the per bin SNR is not the optimal metric for molecular detection.\cite{2025JB_MHRS, 2026_Chen_ErrorBudget} Here molecular detections can be made at high significance even when the SNR in each bin is quite low and is an interesting vein of ongoing investigation.  

\section{Conclusions}
\label{sec: Conclusions}

We have developed a formalism to incorporate wavefront error (WFE), wavefront sensing and control (WFS\&C), and coronagraph sensitivity information into exposure time calculations and presented Error Budget Software (EBS), an open-source tool to implement this work into the existing EXOSIMS ETC to begin performing trade studies quickly as motivated by the rapid progress of the Habitable Worlds Observatory. This tool is publicly available  (\url{https://github.com/chen-pin/ebs}), and it has been benchmarked against other existing ETCs \cite{2025ETCCrossCal}.

We also performed a series initial trade studies with EBS to serve as examples of how EBS can be utilized and to uncover key initial insights. The first shows that the interplay between coronagraph raw contrast and WFE (observatory stability) follows a power law with an exponent of two up until the point that exo-Earths can be imaged in single exposures at which point exposure times are agnostic to additional input WFE or gains from WFS\&C. We then looked at the relationship between energy resolution ($R$) and dark current setting other detector noise terms to 0. For low values of dark current ($\xi < 1\times10^{-6}$) exposure time increases linearly with $R$ and at high values of dark current ($\xi > 1\times10^{-2}$) exposure time increases quadratically with $R$ as is explained analytically in Eq.~\ref{eq: ttsnr_spec_full}. Interestingly this transition happens around current best detector dark current values meaning that large gains in integration time can potentially be achieved for modest improvement in dark current, especially for higher values of $R$. The location of this transition, however, changes as a function of exo-zodi level. For pessimistic exo-zodi scenarios (z $\sim$ 100), the exo-zodi shot noise will dominate over the detector noise in the error budget loosening the requirements on detector noise for our stated observatory and astrophysical assumptions. Finally we performed two MCMC studies to demonstrate EBS's capability to conduct ``global" design explorations and its usefulness in identifying design break points, sensitivities, and trades.      

In order for HWO to achieve its full science potential, trades must be performed that account for the full observatory system, and not just individual science instruments and components. Importantly this includes observatory instabilities that show up as wavefront errors, the mitigating effect of active wavefront sensing and control, and coronagraph sensitivity to these errors at a variety of spatial and temporal scales. We hope that EBS will allow for these key parameters to be more easily incorporated into ongoing trade studies with the Exploratory Analytic Cases (EACs) and will help the HWO TMPO refine the design of the observatory over the next decade. 

\subsection* {Disclosures}
The authors declare that there are no financial interests, commercial affiliations, or other potential conflicts of interest that have influenced the objectivity of this research or the writing of this paper.

\subsection* {Code and Data Availability}
The code used for this work is publicly available at \url{https://github.com/chen-pin/ebs}. EBS makes use of the matplotlib\cite{Huntermatplotlib}, astropy\cite{2022astropy}, numpy\cite{harris2020numpy}, EXOSIMS\cite{2016SPIEDelacroix}, emcee\cite{2013emcee}, and dynesty\cite{dynesty_paper, dynesty_zenodo} packages.

\subsection* {Acknowledgments}
We would like to thank the referees for their helpful comments, which improved the quality of this work.

We acknowledge support by NASA Exoplanet Exploration Program (ExEP), as part of the Coronagraph Technology Roadmap effort. Sarah Steiger acknowledges support from an STScI Postdoctoral Fellowship. The research at the Jet Propulsion Laboratory (JPL), California Institute of Technology, was performed under contract with the National Aeronautics and Space Administration.

\bibliography{report}   
\bibliographystyle{spiejour}   


\noindent\textbf{Sarah Steiger} is an STScI Postdoctoral Research Fellow working with the Russell B. Makidon Optics Laboratory and Community Missions Office. She received her doctorate from the University of California, Santa Barbara in 2023 and is an expert in high-contrast imaging and superconducting detector technologies. Her current focus is in the development of new technologies to enable the exoplanet imaging goals of future flagship space observatories including detectors, coronagraphs, and wavefront sensing and control systems.

\noindent\textbf{Pin Chen} is a planetary scientist at the NASA Jet Propulsion Laboratory, California Institute of Technology (JPL/Caltech). He received his doctorate degree from Caltech in 1999, and he was a NRC Postdoctoral Research Associateship fellow at NIST in Boulder, Colorado, before becoming a JPL scientist. His research currently focuses on exoplanet direct imaging/characterization, atmospheric chemistry, and Early Earth system chemistry. He currently serves as a Deputy Pre-Formulation Scientist for the Habitable Worlds Observatory.

\noindent\textbf{Laurent A. Pueyo} is an associate astronomer at Space Telescope Science Institute (STScI). After receiving his doctorate from Princeton University in 2008, he worked as a NASA fellow at JPL and a Sagan fellow at Johns Hopkins University (JHU). His research focuses on imaging faint planets around nearby stars. He has pioneered optical technologies that allow astronomers to image other planetary systems and has developed data analysis methods now standardly used to study extrasolar planets.



\end{spacing}
\end{document}